\documentclass[12pt]{article}
\let\a=\alpha\let\b=\beta\let\d=\delta
\let\e=\epsilon\let\g=\gamma
\let\th=\theta\let\k=\kappa\let\l=\lambda
\let\m=\mu\let\n=\nu\let\r=\rho
\let\s=\sigma\let\t=\tau
\let\w=\omega\let\y=\psi
\let\z=\zeta\let\G=\Gamma
\let\D=\Delta
\let\vt=\vartheta 
\newcommand{\nn}{\nonumber}
\newcommand{\un}{\underline}
\newcommand{\be}{\begin{equation}}
\newcommand{\ee}{\end{equation}}
\newcommand{\bea}{\begin{eqnarray}}
\newcommand{\eea}{\end{eqnarray}}

\renewcommand{\paragraph}[1]{
\vspace{.8mm}\par\noindent {\sl #1}\\
\vspace{0.2mm} }

\newcommand{\rep}{c}

\newcommand{\ft}[2]{{\textstyle\frac{#1}{#2}}}

\newcommand{\ba}{\left(\begin{array}}
\newcommand{\ea}{\end{array}\right)}

\def\Ggp{G}

\newcommand{\Ca}{{\bf c}}
\newcommand{\Ha}{{\bf h}}
\newcommand{\Ga}{{\bf g}}

\def\G{{\bf G}}

\def\ua{{\underline \alpha}}

\newfont{\Bbb}{msbm10 scaled 1200}     
\newcommand{\mathbb}[1]{\mbox{\Bbb #1}}

\def\IR{{\mathbb R}}

\def\hmu{{\hat{\mu}}}
\def\hnu{{\hat{\nu}}}
\def\halpha{{\hat{\alpha}}}
\def\hbeta{{\hat{\beta}}}
\def\hgamma{{\hat{\gamma}}}
\def\tr{{\rm{tr}}}
\def\cdd{{\cal D}}

 \def\adscft{$AdS$/$CFT$}
\def\G{\Gamma}  

\def\ha{\hat{\a}}
\def\hb{\hat{\b}}
\def\hg{\hat{\g}}
\def\hd{\hat{\d}}

\def\hl{\hat{\l}}
\def\hm{\hat{\m}}
\def\hn{\hat{\n}}
\def\hr{\hat{\r}}
\def\Fab{F_{\ha \hb}}

\def\psl{p  \! \!  /}

\def\cD{{\cal D}}
\def\Dsl{\cD \! \! \! \! /}
\def\dsl{\nabla \! \! \! \! /}
\newcommand\SYM{super-Yang-Mills }
\newcommand\eom{equations of motion }

\def\Gm{\G_{\hm}}

\def\Fmn{F_{\hm \hn}}

\def\Aa{A_{\ha}}

\def\Am{A_{\hm}}
\def\am{a_{\hm}}

\def\um{{\un m}}

\def\adscft{$AdS$/$CFT$ }
\def\adsxs{AdS_5 \times S^5}
\def\Mtt{C_{(10|32)}}
\def\Msx{C_{(10|16)}}
\renewcommand{\un}[1]{{\underline{#1}}}

\def\ua{{\underline a}}

\def\um{{\underline m}}

\newsavebox{\uuunit}
\sbox{\uuunit}
{\setlength{\unitlength}{0.825em}
 \begin{picture}(0.6,0.7)
\thinlines
\put(0,0){\line(1,0){0.5}}
\put(0.15,0){\line(0,1){0.7}}
 \put(0.35,0){\line(0,1){0.8}}
\multiput(0.3,0.8)(-0.04,-0.02){12}{\rule{0.5pt}{0.5pt}}
\end {picture}}

\begin{document}
\begin{titlepage}
\begin{flushright}
\vspace*{-2.2cm}
{\tt CALT-68-2268}\\
{\tt CIT-USC/00-026}\\
{\tt UCB-PTH-00/14}\\

{\tt hep-th/0007104}\\
\end{flushright}
\vskip 1cm

\begin{center}
{\LARGE {\bf  Holography in Superspace}}
\vskip 1.2cm

{ Hirosi Ooguri\footnote{On leave of absence from
the University of California, Berkeley.}, J. Rahmfeld,
 Harlan Robins$^{1}$,
and Jonathan Tannenhauser$^{1}$}\\

\vspace{1cm}

{\small
{\sl California Institute of Technology, Mail Code 452-48,\\
Pasadena, CA 91125}
\vspace{0.1cm}

{\sl and}
\vspace{0.1cm}

{\sl CIT-USC Center for Theoretical Physics, \\University of
Southern California,
Los Angeles, CA 90089
}}

\vspace{0.3cm}

{\tt   ooguri, rahmfeld, hrobins, jetannen@theory.caltech.edu}

\vskip 1cm
\begin{abstract}
The \adscft correspondence identifies the coordinates of the
conformal boundary of anti-de~Sitter space with the coordinates
of the conformal field theory.  We generalize this identification
to theories formulated in superspace.  As an application
of our results, we study a class of Wilson loops in ${\cal N} = 4$ \SYM
theory. A gauge theory computation shows that the expectation values 
of these loops are invariant under a local $\k$-symmetry, 
except at intersections. We identify this with the $\k$-invariance 
of the associated string worldsheets in the corresponding bulk
superspace.

\end{abstract}

\end{center}

\vfill

\end{titlepage}

\section{Introduction}
One of the key ingredients in the \adscft correspondence
\cite{malda} is the identification of the conformal boundary of
anti-de~Sitter space with the space on which the conformal field
theory is defined.   In $AdS$ space with Poincar\'e
coordinates $(x^{\m},y)$ and metric \be
   ds^2 = {1 \over y^2} (dx^2 + dy^2),
\label{adsmetric} \ee the boundary is located at $y=0$, and we
identify the $x^\mu$ as the coordinates of the dual
conformal field theory. Given this identification of coordinates, the 
isometries of $AdS$, restricted to the boundary, are identified with the
conformal symmetries of the boundary.
Consider, for example, the infinitesimal isometry
\bea
\label{babyisometries}
    \delta x^\mu &=& -2(\epsilon \cdot x) x^\mu
                 + \epsilon^\mu ( x^2 + y^2), \nn \\
   \delta y &=& -2 (\epsilon \cdot x)y,
\eea
which preserves the metric (\ref{adsmetric}). As $y
\rightarrow 0$, the coordinate $y$ decouples from the isometry
action, and $\delta x^\mu$ reduces to a special conformal
transformation of the boundary.

Do similar relations hold in bulk-boundary pairs of superspaces? 
One would expect any such relations to involve the fermionic coordinates
in an interesting way. (For example,  
$AdS_5\times S^5$ superspace has 32 fermionic coordinates, but
 there are only 16 fermionic 
coordinates in the boundary theory.)  Bosonic $AdS$ space and 
its conformal boundary can be realized as coset manifolds of 
the same group. This suggests studying the bulk-boundary correspondence
in superspace from the point of view of coset supermanifolds. 
In section 2, we show that, under certain 
conditions, the symmetries of a bulk-boundary pair of coset supermanifolds
coincide in an appropriately defined boundary limit.  We then delve 
into the identification of coordinates and symmetries in the case 
that the bulk supermanifold is the $(10|32)$-dimensional 
$\adsxs$ superspace whose bosonic part is $AdS_5 \times S^5$, 
and the boundary is the conformal superspace 
on which the ${\cal N} = 4$ \SYM theory
is naturally defined\footnote{Our
analysis continues the study of the symmetries of these spaces begun
in \cite{CRRTY}.}.

We next apply the superspace approach 
to the correspondence between Wilson loops
in ${\cal N} = 4$ \SYM theory and string worldsheets in $AdS_5 \times
S^5$ \cite{malda2, reyyee}.  In section 3, we define the Wilson loop
operator $W$ in the superfield formulation of ${\cal N} = 4$ \SYM theory.
If the loop is lightlike (that is, if its
tangent vector in superspace is everywhere null), 
$W$ is invariant under a local $\k$-symmetry, provided
the classical super-Yang-Mills
equations of motion are satisfied.  The $\k$-symmetry in question is
defined in section 3 and  is essentially the usual $\k$-symmetry of a
massless superparticle.   Of course, the
quantity that enters into the \adscft correspondence is not the classical
operator $W$, but rather its quantum expectation value $\langle W
\rangle$.
Replacing classical equations of motion by Schwinger-Dyson equations,
we are able to show that $\langle W \rangle$, to lowest nontrivial
order in an expansion in fermionic superspace coordinates, is $\k$-invariant,
provided the loop is lightlike and smooth and 
contains no self-intersections.  On the other hand, 
we find that the $\kappa$-symmetry is violated when
the loop has intersections.

According to the \adscft correspondence, Wilson loops in the boundary
gauge theory are associated with worldsheets in the bulk
string theory. We employ the approach of \cite{Bergshoeff,Tseytlin}
to study  the Green-Schwarz superstring propagating  
in the 
$AdS_5 \times S^5$ superspace\footnote{Because of the $\kappa$-symmetry,
the approach of \cite{Tseytlin} is difficult to use to describe 
short strings in the bulk, but has been useful
in studying string worldsheets of macroscopic size
ending on Wilson loops on the boundary \cite{Forste,Kinar,dgt}.}. 
The bulk string worldsheets are constrained to end on the Wilson
loop, which lives in the conformal boundary of the $AdS$
space.  In other words, the Wilson loop imposes boundary conditions on
the string worldsheet with which it is associated \cite{malda2, reyyee}.
Using the coordinate identifications worked out in section 2, we study
these boundary conditions in section 4, extending to superspace the
analysis
carried out in \cite{dgo} for bosonic loops. We learn, for example, that
the
requirement that the loop be lightlike, which we arrived at in
section 3 for wholly gauge-theoretic reasons, has an interpretation
in the context of \adscft as a necessary condition for the string
worldsheet to terminate on the boundary of the $AdS$ space.   Finally,
we show that the $\k$-symmetry variation of the Wilson loop in the
gauge theory can be understood as the restriction to the boundary of
the $\k$-symmetry of the associated string worldsheet in $AdS_5
\times S^5$ superspace.

\newpage

\section*{Notation}

\noindent
$\bullet$ Indices

$\hmu, \hnu=0,\dots, 9$ are vector indices in 10 dimensions.

$\mu, \nu = 0, \dots, 3$ are vector indices in 4 dimensions.

$m, n=4, \dots, 9$ are vector indices in 6 dimensions.

\vskip .1in

$\halpha, \hbeta = 1, \dots, 16$ are Weyl spinor indices in 10
dimensions.

$\alpha, \beta = 1, \dots, 4$ are Dirac spinor indices in 4 dimensions.

$a, b= 1, \dots, 4$ are Weyl spinor indices in 6 dimensions.

\noindent
These are indices for general coordinates. The indices
for the local Lorentz frame are obtained by underlining
the corresponding general coordinate indices,
$e.g.$, $\mu \rightarrow \underline{\mu}$.

\vskip .15in
\noindent
$\bullet$ Coordinates

\noindent
The $(10|16)$-dimensional {\sl boundary} superspace has coordinates
\be
z^M =(x^{\hmu}, \lambda^{\halpha})
=(x^\mu, y^m, \lambda^{\halpha}).
\ee
The superspace derivatives are
\bea \partial_{\hmu} &=& {\partial \over \partial x^{\hmu}}, \nn \\
D_{\halpha} &=& {\partial \over \partial \lambda^{\halpha}}
 + (\Gamma^{\hmu} \lambda)^{\halpha} {\partial \over \partial x^{\hmu}},
\nn \\
Q_{\halpha} &=& {\partial \over \partial \lambda^{\halpha}}
 - (\Gamma^{\hmu} \lambda)^{\halpha} {\partial \over \partial x^{\hmu}}, \nn
\\
{\bf D} &=& \lambda^{\halpha} {\partial \over
\partial \lambda^{\halpha}} = \lambda^\halpha D_\halpha.
\label{superder}
\eea

\noindent
The $(10|32)$-dimensional {\sl bulk} superspace has coordinates
\be
Z^{{\bf M}}= (X^{\hmu},  \theta^{\halpha},
\vartheta^{\halpha}) = (X^\mu, Y^m,
\theta^{\halpha}, \vartheta^{\halpha}).
\ee
Sometimes we write $Y^m = (Y, \phi^m)$, where
$Y$ is the radial coordinate in the $AdS$ space and the $\phi^m$'s are
coordinates on the 5-sphere. 





\newpage

\section{Supercosets and Holography}
\setcounter{equation}{0}
It is well known that the isometries of $AdS$ reduce on the boundary to
conformal transformations.  As we saw in the example
(1.2), the radial coordinate
$y$ decouples from the transformations of the coordinates parallel
to the boundary. These parallel coordinates are naturally
identified with the coordinates of the boundary conformal
field theory. Adding
supersymmetry makes the boundary limit analysis more interesting.
Consider, for example,
$(10|32)$-dimensional $AdS_5 \times S^5$ superspace.  In addition to
$y$, half of the fermionic variables must decouple from the isometries
at the boundary: the bulk theory has 32 supersymmetries, whereas the
boundary theory, being a non-gravitational
theory, admits only 16 linear supersymmetries.  
Thus, the 32 fermions of the bulk theory split into two
sets of 16.  One set decouples from the isometries in the boundary
limit, and the other set maps in this limit to the 16 fermionic
coordinates of the conformal field theory.

In this section, we analyze the identification of
coordinates and symmetries from the perspective of coset manifolds.
This point of view is fruitful for the generalization to superspace,
which is our main interest. In section 2.1, we briefly review coset 
techniques.  In section 2.2, we exhibit the boundary reduction
of $AdS_5$ isometries to four-dimensional conformal symmetries,
presenting the ideas from the coset manifold point of view.
Section 2.3 develops a general formalism for comparing the
symmetries of two coset manifolds of the same group.   
Finally, in section 2.4, we
apply our general techniques to the $(10|32)$- and
$(10|16)$-dimensional supercoset manifolds that enter in the
superspace formulation of the $AdS_5/CFT_4$ correspondence.  
We match coordinates and symmetries at the boundary, verify that 
half of the $32$ $AdS$ fermionic coordinates decouple, 
and discuss the relation of the supervielbeins of the two spaces. 


\subsection{A Brief Review of Supercosets}

Let $C = G/H$ be a coset (or supercoset) manifold with 
coordinates $Z^M$,
and let us choose the coset representative $c(Z) \in G$.  The coset  
representative is
defined up to the equivalence relation
\be
\rep(Z)\sim  \rep(Z)h(Z)\, ,
\label{equiv}
\ee
where $h(Z)$ is a local $H$ transformation.
The Lie algebra-valued Cartan 1-form ${\cal L}$ is
defined as
\begin{equation}
{\cal L}(Z) ={\cal L}^A(Z) \Ga_A = dZ^M {\cal L}_M{}^A \Ga_A \equiv
\rep (Z)^{-1} d \rep (Z)\, ,
\label{cartandef}
\end{equation}
where the $\Ga_A$ are the generators of $G$. The Cartan form
${\cal L}$ can be written as
\be
{\cal L}^A \Ga_A = E + \omega = E^{\underline M} \Ca_{\underline M} +
\omega^i \Ha_i\, .
\label{CartanDecompo}
\ee
In this expression, $\Ca_{\underline M}$ and $\Ha_i$ denote the coset
and stability group generators, respectively; their coefficients
$E^{\underline M}$ and $\omega^i$ are the supervielbein and the 
$\Ha$-connection, which is a generalization of the usual spin connection.
The equivalence relation  (\ref{equiv}) induces the identifications 
\bea
E^{\underline M}&\sim& \left(h(Z)^{-1}E h(Z)\right)^{\underline M},
\label{Identi}\\
\omega^i&\sim&  \left(h(Z)^{-1}{\cal L} h(Z)\right)^{i}+h(Z)^{-1} dh(Z) \,
.\nn
\eea
In particular, the vielbein is defined only up to local $H$-transformations.

Under left-multiplication by a constant $g\in\Ggp$, the
coset representative transforms as
\be
\rep(Z)\rightarrow \rep(Z')=g\rep(Z)h(Z)^{-1}\, .
\label{trans}
\ee
The compensating transformation $h(Z)$ ensures that $\rep(Z')$ 
remains in the same gauge slice. 
The infinitesimal form of (\ref{trans}) reads
\be
\d \rep(Z)={\bf g} \rep(Z)-\rep(Z) {\bf h}(Z)\, ,
\label{infTransN}
\ee
where
\bea
g&=&1+{\bf g} ,\nn \\
h&=&1+{\bf h}(Z) \, .
\label{eq:infinitesimal}
\eea

The left-multiplication (\ref{trans}) induces a transformation
of the coordinates $Z^M$ via
\be
\d \rep(Z)= \d Z^M \partial_M \rep(Z) \, .
\ee
These global symmetries leave the vielbein invariant up to the local {\bf
h}-transformation
\be
E^{\underline M}\rightarrow E^{\underline M}+ [{\bf
h}(Z),E]^{\underline M}\, .
\label{htransf}
\ee
If $G$ and $H$ are semisimple bosonic
groups, the
unique and natural line element is invariant under symmetries of this
form. In this case,
these symmetries are isometries in the usual sense.


\subsection{AdS/CFT: The Bosonic Case}

Now let us turn to the bosonic cosets involved in the $AdS_5/CFT_4$
correspondence.
The bulk space $AdS_5$ is the coset manifold $G/H=SO(2,4)/SO(1,4)$, and is 
parametrized by four boundary coordinates $x^\mu$ and the radial coordinate
$y$. 
The symmetries of the boundary $\IR^4$
are made manifest if we realize this space as the coset
$G/H'=SO(2,4)/{\rm Span}(iso(1,3)_K\oplus D)$, which
we refer to as conformal space. 
Here Span($\cdots$) denotes the group generated
by the operators in ($\cdots$).  The stability group $H'$ is not
semi-simple, and 
the $G$-action preserves the natural metric
on $\IR^4$ only up to rescaling. 
The $SO(2,4)$ generators and their conformal 
weights (with respect to the dilatation operator $D$) are listed
in Table \ref{tab:Bgenerators}.   The
classification of generators by coset and stability group is given in
Table \ref{tab:Bcosets}.
\begin{table}[hbtp]
\begin{center}
\begin{tabular}{|c|c|l|} \hline
Operator&Weight&Name\\ \hline\hline
$P_{\m}$&$1$& Conformal Translations\\ \hline
$M_{\m\n}$&$0$& Lorentz Rotations\\ \hline
$D$&$0$&Dilatation\\ \hline
$K_\mu$&$-1$&Special Conformal Transformations\\ \hline
\end{tabular}
\end{center}
\caption{$SO(2,4)$ generators in the conformal basis
\label{tab:Bgenerators}}
\end{table}
\begin{table}[hbtp]
\begin{center}
\begin{tabular}{|c||c|c|} \hline
 &$AdS_5$& $\IR^4$\\ \hline\hline
&&\\
~~$C$~~&$\frac{1}{2}(P_\m+K_\m),~~D$& $P_\m$ \\
&&\\ \hline &&\\
~~$H$~~& $\frac{1}{2}(P_\m-K_\m),
~~M_{\mu\nu}$& $K_{\m},~~M_{\mu\nu}$ \\
&&\\\cline{2-3}
\hline
\end{tabular}
\end{center}
\caption{Coset decompositions of $SO(2,4)$ Generators
\label{tab:Bcosets}}
\end{table}
To relate $AdS_5$ and conformal space,  we first select
coset representatives.  A convenient choice of
representative for the $AdS$ coset is given \cite{Ferrara}  by
\be
\rep_{AdS}(x)=e^{x^\mu P_\mu} y^D\, .
\label{BosRepA}
\ee
The advantage of this choice is that the coset
generators are ordered by weight, which simplifies the form of the
Cartan form and the symmetry transformations.
For the coset representative of the conformal
space $\IR^4$  we choose the standard form
\be
\rep_{cf}(x)=e^{x^\mu P_\mu} \, .
\label{BosRepC}
\ee
Strictly speaking, we should introduce different
symbols $\tilde x$ to denote coordinates for this space,
but as we show in the next subsection, both the vielbeins
in the $x$ directions and the symmetries agree in the limit $y\rightarrow
0$,
so we allow this imprecise notation for better readability.

The $AdS$ coset representative (\ref{BosRepA}) gives rise
to the $AdS$ vielbein \cite{Ferrara}
\be
E^{\underline \mu} = \frac{dx^{\underline \m}}{y}, \qquad E^{\underline
y}= \frac{dy}{y}\, .
\label{AdSvielbein}
\ee
The vielbein $e$ of conformal space is
\be
e^{\underline \mu}= dx^{\underline \mu} ,
\label{Confvielbein}
\ee
which agrees with  the parallel components of (\ref{AdSvielbein})
up to a conformal
rescaling. 
This conformal rescaling is a ``gauge transformation'' of precisely the
form (\ref{htransf}), with ${\bf h} = \ft {1-y}{y} D$.

Let us revisit the well-known correspondence between
bulk isometries and boundary conformal symmetries
from the point of view of the coset construction. 
Consider the special conformal transformation with ${\bf g} = \e^{\mu}
K_{\mu}$.
The action of this symmetry on the coordinates
of conformal space is
\be
\d_{cf} x^{\m} =  x^2 \e^{\m} - 2 (\e \cdot x) x^{\m}\, .
\ee
To calculate the corresponding isometry of the $AdS$ coset, 
we note that the gauge choice (\ref{BosRepA}) requires that 
the ${\bf g}$-action be accompanied by a compensating
${\bf h}$-transformation of the form 
\be
\label{Hform}
{\bf h}_{AdS} = -4 x^{\m} \e^{\n} M_{\m \n} -  y \e^{\m} (P_{\m} - K_{\m}).
\ee
We see that ${\bf h}_{AdS}$ includes a modifying translation, because the
stability group includes the combination $\ft 1 2(P_{\m} - K_{\m})$. From 
here, we can read off
\be
\d_{AdS} x^{\m} =  x^2 \e^{\m} - 2 (\e \cdot x) x^{\m} + y^2 \e^{\m}\, .
\label{AdSKsym}
\ee
Therefore,
\be
\label{deltadelta}
\Delta (\d x^\mu)\equiv \d_{AdS} x^{\m}-\d_{conf} x^{\m} = y^2 \e^\mu\,
,
\ee
which vanishes on the boundary $y \rightarrow 0$.
It is worthwhile to note that the crucial condition for the agreement
between the two variations is
\be
\left.
\left(y^D{\bf h}_{AdS}(x,y) y^{-D}\right)
\right|_{P_\mu}\sim y^2 \rightarrow 0
\, .
\label{CritCase}
\ee

\subsection{The General Picture}

We now generalize the above discussion to arbitrary supercosets.
Consider two cosets $C_1 = G / H_1$ and $C_2 = G
/ H_2$ with the same underlying group $G$, but different stability
groups $H_1$ and $H_2$. Let their coordinates be by $Z^M = (x^m, y^i)$
and $x^m$, respectively (again, this notation anticipates that the
$x^m$ coordinates of both spaces can be identified in a suitable
limit). In this subsection only, $x$ and $y$ can denote  
either bosonic or fermionic coordinates.  
We further suppose that the coset representative of $C_1$ has the form
\be
\label{repfactor}
c_1(x,y)=c_2(x) h_2(x,y),
\ee
where $h_2(x,y)$ is an $(x,y)$-dependent  element of $H_2$ and
$c_2(x)$ is a coset representative of $C_2$.

How are the symmetries of $C_1$ and $C_2$ related?
To begin with,
\be
\label{cosetisom3}
c^{-1} ({\bf g} \, c - c {\bf h}) = \d z^M {\cal L}_M,
\ee
where
\be
\label{Cartanform1}
{\cal L}_M = E_M{}^{\um} {\bf c}_{\um} + \w_M{}^i {\bf h}_i\, .
\ee
Applying (\ref{cosetisom3}) to the coset $C_1$ and making use of the
factorization
(\ref{repfactor}) gives
\be
\label{onehand}
c_2^{-1}({\bf g} c_1 - c_1{\bf h} _1) h_2^{-1} =
\d_1 x^m ({\cal L}_m^{(2)} + (\partial_m h_2) h_2^{-1}) + \d_1 y^i
(\partial_i h_2) h_2^{-1},
\ee
where ${\bf h}_1 = {\bf h}_1({\bf g},x,y)$ is the compensating
transformation for $G$ within
coset 1,
the expressions ${\cal L}_m^{(2)} = c_2^{-1}(x) \partial_m c_2(x)$ are
components
of the Cartan form of $C_2$, and the notation
$\d_1$ reminds us that we are
considering symmetries of $C_1$.  On the other hand,
\bea
\label{otherhand}
c_2^{-1} ({\bf g} c_1 - c_1 {\bf h}_1) h_2^{-1} &=& c_2^{-1} {\bf g} c_2 -
h_2
{\bf h}_1 h_2^{-1} \nn \\ &=& c_2^{-1} {\bf g} c_2 - {\bf h}_2 + {\bf h}_2
- h_2 {\bf h}_1
h_2^{-1} \nn \\ &=& \d_2 x^m {\cal L}_m^{(2)} + {\bf h}_2 - h_2 {\bf h}_1
h_2^{-1}.
\eea
Here ${\bf h}_2 = {\bf h}_2({\bf g},x,y)$ is the compensating
transformation for ${\bf g}$ in
coset 2, and $\d_2 x^m$ is the corresponding infinitesimal
coordinate variation.  From (\ref{onehand}) and (\ref{otherhand}), 
it follows that
\bea
\label{bothhands}
\D(\d x^m) {\cal L}_m^{(2)} &\equiv& (\d_1 x^m - \d_2 x^m){\cal L}_m^{(2)}  
\nn \\ &=& {\bf h}_2 - h_2 {\bf h}_1 h_2^{-1} - 
\d_1 x^m (\partial_m h_2) h_2^{-1} - \d_1 y^i (\partial_i
h_2) h_2^{-1}.
\eea
We now compare the coefficients of the $C_2$ generators on both sides of
(\ref{bothhands}). On the right-hand side, the only term with a
potentially nonzero coefficient is $-h_2 {\bf h}_1 h_2^{-1}$; the other
terms lie exclusively in the Lie algebra of the stability group
$H_2$.  Consequently,
\be
\D (\d x^m) E^{(2)}_m{}^{\um} = \left. - \left( h_2 {\bf h}_1 h_2^{-1}
\right)
\right|_{C^{(2)}_{\um}}.
\ee
The vielbein $E^{(2)}$ is invertible and is independent of
$y$. The symmetries of $C_1$ and $C_2$ agree, $i.e.$, 
$\Delta (\delta x^m)=0$, if 
\be
\label{criterion}
 \left. \left(
h_2 {\bf h}_1 h_2^{-1} \right) \right|_{C^{(2)}_{\um}} =0,
\ee
at some value of $y$ ($y=0$, in the $AdS$ example). In this case,
we say $C_2$ is the boundary limit of $C_1$, located
at $y=0$. This condition generalizes (\ref{CritCase}).

\subsection{AdS/CFT: The Supersymmetric Case}
Both $AdS_5\times S^5$ superspace and conformal superspace
are supercosets of $G=SU(2,2|4)$, but with different stability groups.
Table \ref{tab:generators} lists the generators of  $SU(2,2|4)$ with their 
weights under the dilatation
operator $D$.
\begin{table}[hbtp]
\begin{center}
\begin{tabular}{|c|c|l|} \hline
Operator&Weight&Name\\ \hline\hline
$P_\mu$&1& Conformal Translations\\ \hline
$Q$&$1/2$& Global Supersymmetries\\ \hline
$M_{\mu\nu}$&0& Lorentz Rotations\\ \hline
$D$&$0$&Dilatation\\ \hline
$U^{~i}_j=(\widetilde{M}_{m'n'}, \widetilde{P}_{m'})$
&$0$&$SU(4)$ Rotations of $S^5$\\ \hline
$S$&$ - 1/2$&Special Supersymmetries\\ \hline
$K_{\mu}$&$- 1$&Special Conformal Transformations\\ \hline
\end{tabular}
\end{center}
\caption{\small{$SU(2,2|4)$ generators in the superconformal basis.
The
generators of $SU(4)$ rotations may be written as $U_i^j = 
2\widetilde{P}_{m'}
(\widetilde{\G}^{m' 6})_i^j + \widetilde{M}_{m' n'} (\widetilde{\G}^{m' n'})_i^j$, 
where the $\widetilde{P}_{m'}$
and $\widetilde{M}_{m' n'}$ 
($m',n' = 1, \dots, 5$) are generators of translations
and rotations on $S^5$, and the $\widetilde{\G}$'s are the $4 \times 4$
chiral blocks of the $SO(6)$ Dirac matrices in the chiral basis.}}
\label{tab:generators}
\end{table}
The $AdS_5 \times S^5$ supercoset
\be
 C_{(10|32)}=\frac{SU(2,2|4)}{SO(1,4)\times SO(5)}\, 
\ee
is $(10|32)$-dimensional, as the stability group contains no fermionic
generators. On the other hand, the
conformal superspace
\be
 C_{(10|16)}=\frac{SU(2,2|4)}{{\rm Span}\left(iso(1,3)_K\oplus so(5)\oplus
 S\right)}\, ,
\ee
is $(10|16)$-dimensional.  
This space is an extension of the
$(4|16)$-dimensional superspace
\be
C_{(4|16)}=\frac{SU(2,2|4)}{{\rm
Span}\left(iso(1,3)_K\oplus D \oplus  so(6)\oplus
 S\right)} .
\ee
In Table \ref{tab:cosets} we give the division of the $SU(2,2|4)$ generators 
into coset and stability group generators for each coset space.
\begin{table}[hbtp]
\begin{center}
\begin{tabular}{|c||c|c|c|} \hline
&$C_{(10|32)}$&$C_{(10|16)}$&$C_{(4|16)}$\\ \hline\hline
&&&\\
$C$&$\frac{1}{2}(P_\mu+K_\mu),\widetilde{P}_{m'},D, Q, S$& $P_\mu,~~
\widetilde{P}_{m'}, ~~ D, ~~ Q$ & $P_\mu, ~~~~~~Q$ \\ 
 & & & \\ \hline
&&&\\
$H$& $\frac{1}{2}(P_\mu-K_\mu), M_{\mu\nu},
\widetilde{M}_{m'n'} $ & $K_{\mu}, M_{\mu\nu},
\widetilde{M}_{m'n'}, S$& $K_{\mu}, M_{\mu\nu}, \widetilde{P}_{m'},
\widetilde{M}_{m'n'}, D, S$  \\
&&& \\\hline
\end{tabular}
\end{center}
\caption{Coset decompositions of $SU(2,2|4)$ Generators
\label{tab:cosets}}
\end{table}

\subsubsection{The Supergeometry at the Boundary}

A representative of the $AdS_5 \times S^5$ supercoset that is convenient
for our purpose is given by
\bea
\label{Rep3}
c_{(10|32)} (x,Y,\phi,\theta,\vartheta)&=&e^{x^\m P_\m}
 e^{\bar Q \theta + \bar \theta Q} e^{\bar S \vartheta +\bar \vartheta
S} u(\phi) Y^D \\
 & =& e^{x^\m P_\m} e^{\bar Q \theta + \bar \theta Q}u(\phi) Y^D \times
{Y^{-D}u(\phi)^{-1}e^{\bar S \vartheta +\bar \vartheta S} u(\phi) Y^D}
\nn \\
& =& e^{x^\m P_\m} e^{\left(\bar Q^\a_a\theta_\a^a+\bar \theta^\a_a
Q_\a^a\right)}u(\phi) Y^D
\times e^{\sqrt{Y}\left(\bar S^\a_a\vartheta_\a^b
u_b{}^a+(u^{-1})_a{}^b \bar \vartheta^\a_b
 S_\a^a\right)}\, \nn .
\eea
The matrices $u_a{}^b(\phi)$ are coset representatives 
of the $SO(6)/SO(5)$ subcoset.
The first factor in the last line of (\ref{Rep3})
is the coset representative of the 
$(10|16)$-dimensional conformal superspace,
\be
\rep_{(10|16)}(x,Y, \phi, \theta)=e^{x^\m P_\m}
  e^{\left(\bar Q^\a_a\theta_\a^a+\bar \theta^\a_a
Q_\a^a\right)}u(\phi) Y^D\, ,
\
\label{TenSixteenRep}
\ee
and the second factor is in the stability group of ${\cal
C}_{(10|16)}$, since it contains generators of negative weights only.  
Hence (\ref{Rep3}) is of the form (\ref{repfactor}),
\be
\rep_{(10|32)} = \rep_{(10|16)}(x,Y, \phi, \theta) \times
h_{(10|16)}(x,\theta,\vartheta,Y,\phi) \, .
\label{3216Relation}
\ee
By construction, then, the supervielbeins of
$C_{(10|16)}$ are related to those of $C_{(10|32)}$ by a
coordinate-dependent gauge transformation, as in (\ref{Identi}).  We now
calculate these supervielbeins and exhibit this gauge transformation
explicitly.  We will use the results in section 4.

The Cartan form ${\cal L}^{(10|16)}$ of conformal superspace
decomposes as
\be
{\cal L}^{(10|16)}=c_{(10|16)}^{-1}dc_{(10|16)}=e^{\underline \mu}
P_{\underline \mu} +e^{Y} D+
e^{m'} \widetilde{P}_{m'}
+\bar e^\a_a Q_\a^a+\bar Q^\a_a e_\a^a+\omega^i {\bf h}_i \, ,
\ee
where $\omega^i$ is the ${\bf h}$-connection. 
The supervielbein is  determined to be
\bea
e^{\underline \mu}&=&\frac{1}{Y}\left[dx^{\underline \mu}
+\frac{1}{2}\left(d\bar \theta
\gamma^{\underline \mu}\theta
-\bar \theta \gamma^{\underline \mu} d\theta \right)\right],  \nn \\
e^{\underline Y}&=&\frac{dY}{Y}, \nn    \\
e^{\underline m}&=& e^{\underline m'}(\phi), \label{ConfViel}    \\
e_{\underline \a}^{\underline a}&=&Y^{-\frac{1}{2}}
d\theta_{\underline \a}^{\underline b} 
u(\phi)_{\underline b}{}^{\underline a},\nn \\
\bar e^{\underline \a}_{\underline a}&=&
Y^{-\frac{1}{2}}u^{-1}(\phi)_{\underline a}{}^{\underline b}
d\bar \theta^{\underline \a}_{\underline b}\, . \nn
\eea
The connection has components only in the sphere directions,
\be
\omega=\omega^{m'n'}(\phi)\widetilde{M}_{m'n'}\, .
\ee

Given the coset representative (\ref{Rep3}), it is straightforward to
calculate the Cartan form ${\cal L}^{(10|32)}$
of the $AdS_5\times S^5$ superspace.  
Since the coset representative $\rep$ is of the
form (\ref{Rep3}), 
the relation between the ${\cal L}^{(10|32)}$ and ${\cal L}^{(10|16)}$ is
\be
{\cal L}^{(10|32)}=h_{(10|16)}^{-1}{\cal L}^{(10|16)} h_{(10|16)}
+h_{(10|16)}^{-1} d h_{(10|16)}\, . \label{16To32}
\ee
The Cartan form ${\cal L}^{(10|32)}$ 
splits under the decomposition of Table
\ref{tab:cosets}
into vielbein and connection terms as
\bea
{\cal L}^{(10|32)}&=&\frac{1}{2}E^{\underline \m} (P_{\underline
\m}+K_{\underline \m}) +
E^{\underline Y} D + E^{\underline m'}(\phi) \tilde P_{\underline m'} \nn \\
& & +\bar E^Q_\ua Q^\ua+\bar Q_\ua E_Q^\ua+\bar E^S_\ua S^\ua+\bar
S_\ua E_S^\ua \nn \\
&& +
\frac{1}{2}\omega^{\underline \m}
(P_{\underline \m}-K_{\underline \m})+ \cdots\, ,
\eea
where the omitted terms contain additional connection components. Using
(\ref{16To32}), we find, to lowest order in $Y$,
\bea
E^{\underline \mu}&=& \frac{1}{Y}\left[dx^{\underline \mu}
+\frac{1}{2}\left(d\bar
\theta
\gamma^{\underline \mu}\theta -\bar \theta \gamma^{\underline \mu} d\theta
\right)\right] +O(Y)
= e^{\underline \mu}+O(Y)\sim  Y^{-1}, \nn \\
E^{\underline Y}&=&\frac{dY}{Y} + O(1)=e^{\underline Y}
+ O(1)\sim Y^{-1} ,\nn \\
E^{\underline m'}&=& e^{\underline m'}
+ O(y)\sim O(1), \label{BoundVielbein} \\
 E_Q^\ua&=& e^{\ua}-Y^{\frac{1}{2}}e^{\underline \mu}\gamma_{\underline
\mu}\vartheta^b u(\phi)_b{}^\ua\sim  Y^{-1/2},\nn
\\
\bar E^Q_\ua&=& \bar e_\ua+Y^{\frac{1}{2}}
u^{-1}(\phi)_\ua{}^b e^{\underline \mu} \bar
\vartheta_b\gamma_{\underline \mu}\sim  Y^{-1/2}\nn \, .
\eea
Note that the bosonic vielbein, restricted to the boundary, is
precisely that of $C_{(10|16)}$. The fermionic
components receive corrections proportional to $\vt$.
Even so, $e^Q$ is gauge-equivalent
to $E^Q$ via a coordinate-dependent ${\bf h}$-transformation of the
form
\be
E^{\underline M} = e^{\underline M} + \left[\sqrt{Y}\left(\bar
S^\a_a\vartheta_\a^b
u_b{}^a+(u^{-1})_a{}^b \bar \vartheta^\a_b
S_\a^a\right),e ~ \right]^{\underline M}\, .
\ee

\subsubsection{From $AdS$ Superisometries to Superconformal Symmetries}

With the tools and experience acquired in the previous subsections, we
 are now ready to prove the main result of this section: that the
 superisometries of the $AdS_5 \times S^5$ superspace
\be
 C_{(10|32)} = {SU(2,2|4)\over SO(1,4) \times SO(5)}
\ee
 reduce near the boundary $Y =0$ of $AdS_5$ to the
 superconformal  transformations of the boundary conformal superspace
\be
C_{(10|16)} = {SU(2,2|4)\over{\rm Span}\left(iso(1,3)_K\oplus so(5)\oplus
 S\right)}.
\ee
We choose the coset representatives (\ref{Rep3}) and
 (\ref{TenSixteenRep}), whose relation is given by
 (\ref{3216Relation}) and (\ref{Rep3}).

Our proof proceeds in two steps.  First, given a superisometry
generator ${\bf g}$ in the Lie superalgebra of $SU(2,2|4)$, we find
the compensating transformation ${\bf h}_{(10|32)}$ in the Lie algebra of
the
stability group $SO(1,4) \times SO(5)$.  In practice, we will not
compute ${\bf h}_{(10|32)}$ in all its glory, but we will extract the
properties that we will need---in particular, the
$Y$-dependence. Given the requisite information about ${\bf h}_{(10|32)}$,
our
second step will be to show that $\left.
\left(h_{(10|16)} {\bf h}_{(10|32)} h_{(10|16)}^{-1} \right)
\right|_{C_{(10|16)}}$ approaches
zero in the limit $Y \to 0$.  This implies via (\ref{criterion}) that the
$C_{(10|32)}$ superisometries and the $C_{(10|16)}$ superconformal
transformations agree on the boundary.

Let us recall from Table 2 that the stability group $H_{(10|32)} \equiv
SO(1,4) \times SO(5)$ is generated by $\ft 1 2 (P_{\m} - K_{\m})$,
$M_{\m \n}$, and $U^H_i{}^j$, where $U^H_i{}^j$ denotes the restriction of 
$U_i^j$ to the generators of the stability group.  The
generators  $K_{\m}$, $M_{\m \n}$ and $U^H_i{}^j$ are shared by
$H_{(10|16)} \equiv {{\rm Span}(iso(1,3|4)_K \oplus so(5)\oplus S)}$.  Therefore,
only the term in ${\bf h}_{(10|32)}$ proportional to the generator  $\ft 1
2 (P_{\m} -
K_{\m})$, and only its $P_{\m}$ part, at that, can contribute to
$\left. \left(h_{(10|16)}{\bf h} _{(10|32)} h_{(10|16)}^{-1} \right)
\right|_{C_{(10|16)}}$: the other terms are projected out in
restricting to the $C_{(10|16)}$ coset generators.  We are thus relieved
of the burden of calculating ${\bf h}_{(10|32)}$ in its entirety.

We would like to determine the $Y$-dependence of the $P_{\m}$ part of
${\bf h}_{(10|32)}$, which, by the preceding remarks, is the same as the
$Y$-dependence of its $\ft 1 2 (P_{\m} - K_{\m})$ piece.  For this
purpose, it is convenient to factor $G_{(10|32)}$ as the product of a term
built from generators of weight $>0$ and a term built from generators
of weight $\le 0$,
\be
c_{(10|32)} = g_+ g_-, \qquad \quad g_+ = e^{x^{\m} P_{\m}} e^{\bar Q
\theta + \bar \theta Q}, \quad g_- = e^{\bar S \vartheta + \bar
\vartheta S} u(\phi) Y^D.
\ee
The coset superisometry generated by ${\bf g}$ satisfies
\be
\d c_{(10|32)} = {\bf g} g_+ g_- -  g_+ g_-  {\bf h}_{(10|32)}.
\ee
Since $g_+$ is nothing but the coset representative of the space
$C_{(4|16)}$, we may likewise write
${\bf g} g_+ = \d_+ g_+ -
g_+ {\bf h}_+$, where $\d_+ g_+$ is the infinitesimal variation of $g_+$
generated by ${\bf g}$, and ${\bf h}_+$ is chosen to guarantee that $\d_+
g_+$
lies in $C_{(4|16)}$.  As we see from Table 4, ${\bf h}_+$
is a linear combination of generators ${\bf g}_i$ of weights $w_i \le 0$,
\be
{\bf h}_+ = \sum_{i: w_i \le 0} a_i{\bf g} _i\, .
\ee
Now, by the Baker-Campbell-Hausdorff formula,
\be
g_+ \left( \sum_{i: w_i \le 0} a_i {\bf g}_i \right) g_- = g_+
g_- \left( \sum_{i: w_i \le 0} b_i Y^{-w_i} {\bf g}_i \right),
\ee
for some new ($Y$-independent) coefficients $b_i$.  Therefore,
\be
\tilde \d c_{(10|32)} \equiv \d c_{(10|32)} -  (\d_+ g_+) g_- =
 g_+ g_- \left( -\sum_{i: w_i \le 0} b_i Y^{-w_i} {\bf g}_i  - {\bf
h}_{(10|32)} \right).
\ee
By construction, $\tilde \d c_{(10|32)}$ is a variation such that
$c_{(10|32)} + \tilde \d c_{(10|32)} \in C_{(10|32)}$.  
We can see this because  $g_+$ is only a function
of $x,\theta,$ and $\bar{\theta}$ only;
it does not depend on the other coordinates of  $C_{(10|32)}$.
The transformation
${\bf h}_{(10|32)}$ is chosen to compensate those terms in $g_+
g_- \left( \sum_{i: w_i \le 0} b_i Y^{-w_i} {\bf g}_i \right)$ which
pull $c_{(10|32)}$ away from the coset. Some of these
terms may be proportional to $K_\mu$.
  Since $K_{\m}$ has weight
-1, the coefficient of $\left. {\bf h}_{(10|32)} \right|_{K_\m} = \left.
{\bf h}_{(10|32)}
\right|_{P_\m - K_\m} =  \left. {\bf h}_{(10|32)} \right|_{P_\m}$ is
proportional to $Y$.

Having established that $\left. {\bf h}_{(10|32)}\right|_{P_{\m}} \sim Y
P_{\m}$, it remains to calculate the behavior of $\left.
\left(h_{(10|16)} {\bf h}_{(10|32)} h_{(10|16)}^{-1} \right)
\right|_{C_{(10|16)}}$ near $Y =
0$.  Straightforward manipulations give
\bea
h_{(10|16)} \left. {\bf h}_{(10|32)}\right|_{P_{\m}} h_{(10|16)}^{-1} &=&
Y^{-D} u^{-1}(\phi) e
^{\bar S \vartheta +\bar \vartheta S} u(\phi)
 Y^D  (Y P_{\m})   Y^{-D} u^{-1} (\phi) e^{-(\bar S \vartheta +\bar
\vartheta
 S)} u(\phi)  Y^D \nn \\ &=&    Y^{-D} \left( u^{-1}(\phi) e^{\bar S
 \vartheta +\bar \vartheta S} u(\phi)(Y^2 P_{\m})u^{-1} (\phi)
 e^{-(\bar S \vartheta +\bar \vartheta S)} u(\phi) \right) Y^D \nn \\
 &=& Y^{-D} (Y^2 {\cal O}) Y^D,
\eea
where the operator ${\cal O}$ is a sum of ($Y$-independent) operators
of weight 1 or lower.
Therefore, $Y^{-D} (Y^2 {\cal O}) Y^D$ contains terms of the form
$Y^{2 - w_i} {\cal O}_i$, with $w_i \le 1$.  The resulting expression
carries a positive power of $Y$, and therefore vanishes at the
boundary $Y = 0$.  By (\ref{criterion}), this establishes the equality
of the $C_{(10|32)}$ superisometries and the $C_{(10|16)}$
superconformal transformations at the boundary.

\newpage

\section{Kappa Symmetry of the Supersymmetric Wilson Loop}
\setcounter{equation}{0}
In the last section, we studied the relation between the
$\adsxs$ superspace $\Mtt$ and the conformal superspace
$\Msx$.  We discovered that, just as in the bosonic case, the
boundary coordinates of $\Msx$ can be identified with a subset
of the bulk coordinates of $\Mtt$.  In the next two sections,
we explore implications of this geometric fact
for Wilson loops in the super-Yang-Mills theory on the boundary.
In particular, we use the identification of coordinates between 
the two superspaces 
to show that the $\k$-symmetry of string worldsheets in type IIB 
string theory on an $\adsxs$ background coincides with the
$\k$-symmetry of the Wilson loops on which the string worldsheets
terminate.

The Green-Schwarz string in the bulk naturally propagates in the 
superspace $\Mtt$ \cite{Tseytlin}.  On the other hand, the 
role of $\Msx$ is evident if we view 
${\cal N}=4$ super-Yang-Mills theory as being obtained by
dimensional reduction from  ${\cal N} = 1$ \SYM theory in ten dimensions.
The Wilson loop operator of this theory is defined as
\bea
  W &=& {1 \over N}\tr P \exp \left( \oint d \t \, \dot z^M e_M^{\un M}
A_{\un M}
\right) \nn \\
    &=& {1 \over N} \tr P \exp \left( \oint d \t \, (A_{\un \hmu}(z(\t))
p^{\un \hmu} + A_{ \un \halpha}(z(\t))
    \dot \lambda^{\un \halpha}(\t)) \right),
\label{10dwilson}
\eea
where the trace is taken in the fundamental of $SU(N)$.  In this
expression, $A_{\un M} = (A_{\un \hmu}, A_{\un \halpha})$ is the
gauge superfield, written in the local Lorentz frame, and
\be
p^{\un \hmu} = \dot x^{\un \hmu}  +  \ft 1 2 \dot \lambda \Gamma^{\un
\hmu}
\lambda  - \ft 1 2 \lambda \Gamma^{\un \hmu} \dot \lambda ,
\label{10dmomentum}
\ee
where $\dot{} = d/d\tau$.
By taking the connection $A_{\underline M}$ to be
independent of the six extra coordinates $y^m$, we
obtain a Wilson loop in the four-dimensional 
${\cal N} = 4$ theory. 
For the remainder of this section, we will not underline 
tangent space indices ($i.e.$, $\underline{\mu} \rightarrow \mu$),
because we will be working entirely in tangent
space.

The gauge superfield $(A_\hmu, A_\halpha)$ defines the covariant
derivatives
\be
  {\cal D}_\hmu = \partial_\hmu + A_\hmu,~~~
{\cal D}_\halpha = D_\halpha + A_\halpha
\ee
on the superspace. The field strengths are defined by
\bea
F_{\ha \b}  &=& \{{\cal D}_{\ha}, {\cal D}_{\hb} \} - 2
\G^{\hm}_{\ha \hb} {\cal D}_{\hm}, \nn \\
F_{\hm \ha} &=& [{\cal D}_{\hm}, {\cal D}_{\ha}], \nn \\
F_{\hmu \hnu}  &=& [{\cal D}_{\hmu}, {\cal D}_{\hnu}]. 
\eea
We also use the spinor superfield defined
by
\be
\label{spinorfield}
\Psi^\halpha = {1 \over 10} \Gamma_{\hmu\hbeta}^{\halpha}
F^{\hmu\hbeta}. 
\ee
In an expansion in fermionic coordinates,
the leading terms of the superfields are given by the physical gauge 
field $a_\hmu(x)$ and the gaugino $\psi^\halpha(x)$,
\be
  A_\hmu(z) = a_\hmu(x) + O(\lambda), ~~~
 \Psi^\halpha(z) = \psi^\halpha + O(\lambda), \ee
and the subleading terms in the expansion contain auxiliary
fields. We will discuss shortly how to remove
these auxiliary fields. After dimensional reduction to four dimensions,
the ten-dimensional gauge field $a_\hmu$ and the gaugino
$\psi^\halpha$ decompose as $a_\hmu = (a_\mu, \varphi_m)$
and $\psi^\halpha = \psi^{\alpha a}$. 

Strictly speaking, the ten-dimensional ${\cal N}=1$ superspace 
is not identical to the conformal
superspace $C_{(10|16)}$.  The two spaces differ in the
action of global supersymmetry on the coordinates $y^m$.  
If $\e$ is
the infinitesimal supersymmetry parameter, then
\be
   \delta_\epsilon y^m = 0
\ee
in $\Msx$ \cite{CRRTY}, whereas
\be
   \delta_\epsilon y^m = \epsilon~ \Gamma^m \lambda
\label{difference}
\ee
in the ${\cal N} = 1$ superspace. For our purposes, this difference is
not significant. After dimensional
reduction to four dimensions, the Wilson loop
(\ref{10dwilson}) depends on $y^m(\t)$ only through
$\dot y^m$ in $p^m$. Therefore, if we redefine $p^\hmu$ as
\bea
    p^\mu &=& \dot x^\mu + \ft 1 2 \dot \lambda \Gamma^\mu \lambda -
\ft 1 2 \l \G^\m \dot \l, \nn \\
    p^m &=& \dot y^m ,
\label{redefinedmomentum}
\eea
it is consistent to set $\delta_\epsilon y^m = 0$, as
in $C_{(10|16)}$. In this section, when we study the properties
of Wilson loops in \SYM theory, we will work in the ten-dimensional ${\cal
N} =1$ superspace, since it simplifies our computation.
In section 4, when we match the $\k$-symmetries of  the Wilson loop and
the string worldsheet, we will make the redefinition
(\ref{redefinedmomentum}),
and regard $z(\t)$ as a loop in $C_{(10|16)}$.

The purpose of this section is to study the $\k$-invariance of the
Wilson loop operator in the gauge theory.  In section 3.1, we define
$\k$-symmetry and prove
that, if the loop is lightlike, $W$ is classically
$\k$-invariant.  Then in section 3.2 we develop the technology
needed for the quantum-mechanical version of the same result:
that under the same hypotheses, the expectation value $\langle W
\rangle$ is $\k$-invariant, to lowest order in a $\l$-expansion.  The
proof is provided in section 3.3, where we also comment on
the case of loops with intersections.

\subsection{Classical Kappa Invariance}

We would like to find a $\k$-symmetry under which the Wilson loop
(\ref{10dwilson}) is invariant.  A natural proposal for the
$\k$-variations of the coordinates is
\bea
  \delta_\kappa x^\hmu &=& -\lambda \Gamma^\hmu \delta_\kappa \lambda,
\nn \\
  \delta_\kappa \lambda & = & \psl \kappa.
\label{firstkappa}
\eea
These are the usual $\k$-invariances of the action of a
superparticle.  Acting with (\ref{firstkappa}) on
(\ref{10dwilson}) yields
\be
\label{deltaloop}
  \delta_\kappa W = - \ft 1 N
  \tr P \oint d\tau \left( p^\hmu  \delta_\kappa \lambda^\halpha
F_{\hmu\halpha}
   + \delta_\kappa \lambda^\halpha \dot \lambda^\hbeta F_{\halpha\hbeta}
\right) \exp\left( \oint d\tau'A_\hmu p^\hmu + A_\halpha \dot \lambda^\halpha
\right) .
\ee

The Wilson loop is $\kappa$-invariant, $\delta_\kappa W=0$,
if 
\be
F_{\halpha \hbeta}= 0,
\label{theconstraintagain}
\ee
and
\be
p^2 = 0
\ee
at every point on the loop. (Note that $p^\hmu(\tau)$ need not be
constant in $\tau$.)  
First of all, if $F_{\ha \hb} = 0$, the term $\delta_\kappa
\lambda^\halpha \dot \lambda^\hbeta F_{\halpha\hbeta}$ in (\ref{deltaloop})
is manifestly zero. Moreover, if  $p^2=0$, the term 
$p^\hmu  \delta_\kappa \lambda^\halpha
F_{\hmu\halpha}$ also vanishes.  To see this, we
start  with the Jacobi identity
\be
[ \{ {\cal D}_{\halpha}, {\cal D}_{\hbeta} \},
{\cal D}_{\hgamma}] + ({\rm cyclic}) = 0.
\ee
Substituting in the definition  of $F_{\ha
\hb}$ and employing the Dirac matrix identities listed in Appendix A.3
gives
\be
   F_{\hmu\halpha}
 = (10 \Gamma_\hmu \Psi)_\halpha
   -{1\over 40} \Gamma_\hmu^{\hbeta\hgamma}\left(
        \cdd_{\halpha} F_{\hbeta\hgamma}
 + ({\rm cyclic}) \right),
\label{bianchi}
\ee
where $\Psi^{\hat \a}$ is as in (\ref{spinorfield}).
The second term in the right-hand side of (\ref{bianchi})
is zero, as $F_{\ha \hb} = 0$.  The first term vanishes
when multiplied by $p^\hmu \delta_\kappa \lambda^\halpha$,
\be
  p^\hmu \delta_\kappa \lambda^\halpha F_{\hmu\halpha}
 = - 10 p^\hmu \delta_\kappa \lambda \Gamma^\hmu \Psi
= - 10 \kappa \psl \psl \Psi = 0 ,
\ee
because $p^2=0$.
Thus we have found that a Wilson loop $W$ obeying
$p^2=0$ everywhere is $\kappa$-invariant, provided the connection
superfield obeys the condition $F_{\ha \hb} = 0$.

What is the meaning of the condition $F_{\ha \hb} = 0$?  As is
typical in superspace formulations of gauge theories, the gauge
superfield $A_M$ contains many auxiliary fields.  To eliminate these
fields in favor of the physical gauge field $a_{\hmu}(x)$ and
gaugino $\psi^{\halpha}(x)$, we must impose a constraint.   The
correct constraint is precisely \cite{witten1}
\be
\label{Fab}
F_{\ha \hb} = 0.
\ee
Moreover, the ${\cal N} = 4$ theory in four dimensions 
has the special property that the
constraint (\ref{Fab}) not only eliminates the auxiliary fields, but
also imposes the equations of motion
\bea
   & & \nabla^{\hmu} f_{\hmu\hnu} + {1 \over 2}
\Gamma_{\hnu  \halpha \hbeta} \{ \psi^{\halpha},
\psi^{\hbeta} \} = 0, \nn \\
& & \dsl ~\psi = 0 
\label{ymeom}
\eea
of the gauge field and the gaugino \cite{witten1, witten2}
($\nabla_\hmu = D_{\hmu} + a_\hmu$;
$f_{\hmu\hnu} = [\nabla_\hmu, \nabla_\hnu ]$). The converse is also
true: if the equations of motion are satisfied, then
the constraint is automatically enforced.  We will see this explicitly
in the next subsection.  This property is an inconvenience if one
wants an off-shell supersymmetric formulation of the ${\cal N}=4$
gauge theory, but for us, it turns out to be a blessing.

We may gain insight into the condition $p^2 = 0$ by recalling that our
$\k$-symmetry variations are those of a massless superparticle.
In \cite{Rocek}, the action of a massless superparticle minimally coupled 
to a background $U(1)$ gauge
superfield was given. This action was shown to be $\k$-invariant
if the gauge fields obey the equations of motion. If we use $p^2=0$,
the kinetic term $\ft 1 2 p^2$ vanishes, 
and what remains is just the exponent of the Wilson loop 
(\ref{10dwilson}).  It is therefore
reasonable that $W$ should be $\k$-invariant if $p^2=0$ and 
the gauge fields are on-shell.

Another interpretation of the condition $p^2 = 0$ was presented in
\cite{witten1,witten2}, where it was shown that the constraint $F_{\ha \hb} =
0$ is equivalent to the condition that the gauge superfield be integrable
on $(1|8)$-dimensional lightlike lines in superspace, with the
8 fermionic dimensions provided by the $\kappa$-transformation.  
Our application to Wilson loops relaxes the requirement that 
$p^{\hm}(\tau)$ be constant.

\subsection{Elimination of Auxiliary Fields}

Equations of motion in a classical field theory lead
to Schwinger-Dyson equations in its quantum counterpart.
Since the \SYM \eom imply
$\delta_\kappa W = 0$ for a lightlike loop, we can
use
the associated Schwinger-Dyson equation to examine whether
the vacuum expectation value $\langle W \rangle$
of the loop remains $\kappa$-invariant in the quantum
theory. To compute $\delta_\kappa
\langle W \rangle$ in practice, we must eliminate the auxiliary fields
and express $W$ in terms of the physical gauge field $a_\hmu$
and the gaugino $\psi$ alone.
A systematic procedure for doing this was
introduced in \cite{hs}.

As we have noted, the constraint $\Fab = 0$ not only eliminates all
auxiliary fields in favor of physical fields $\am$ and $\y$, but also
imposes the equations of motion, and only the equations of motion, on
the
physical fields.  The procedure developed in \cite{hs} separates these
two aspects of the constraint. Let us first discuss
the elimination of the auxiliary fields.  By combining the constraint
(\ref{theconstraintagain}) with the Bianchi identities of the gauge
theory, it is possible to derive the relations \cite{witten2}
\bea
&& F_{\halpha\hbeta} = 0, \nn \\
&& F_{\hmu\halpha} - (\Gamma_\hmu \Psi)_\halpha = 0, \nn \\
&&   {\cal D}_\halpha \Psi^{\hbeta} -
  {1 \over 2} \Gamma^{\hmu\hnu}_{\halpha}{}^{\hbeta} F_{\hmu\hnu}
  = 0, \nn \\
&& {\cal D}_\halpha F_{\hmu\hnu}
   + {\cal D}_{[\hmu,} (\Gamma_{\hnu]} \Psi)_\halpha = 0 .
\label{impliedbyconstraint}
\eea
Now we fix the fermionic gauge invariance.  Following \cite{hs},
we adopt the gauge-fixing condition
\be
   \lambda^\halpha A_\halpha(x,\lambda) = 0.
\label{dgauge}
\ee
We define the operator
\be
{\bf D} = \lambda^\halpha \cD_\halpha
=\lambda^\halpha \partial_{\lambda^\halpha},
\ee
where the second equality is a consequence of the gauge-fixing condition
(\ref{dgauge}).
Multiplying the relations (\ref{impliedbyconstraint}) by $\l^{\ha}$
and using the gauge-fixing condition leads to the ${\bf D}$-{\sl
recursion relations}
\bea
(1 + {\bf D}) \Aa &=& 2 (\Gamma^\hmu\lambda)_\halpha \Am, \nn \\
{\bf D} \Am  &=& -\lambda \Gm \Psi, \nn \\
{\bf D} \Psi^\halpha &=& \ft 1 2 (\lambda \Gamma^{\hmu\hnu})^\halpha \Fmn,
\nn \\
{\bf D} \Fmn &=& \lambda \Gamma_{[\hmu,} \cdd_{\hnu]} \Psi.
\label{drecursion}
\eea
For example, the constraint $F_{\ha \hb} = 0$ gives rise to the
first equation in (\ref{drecursion}), since
\bea
  0 &=& \lambda^{\halpha} F_{\halpha\hbeta} \nn \\
    &=& \lambda^{\halpha} \{\cdd_\halpha, \cdd_\hbeta \}
       - 2(\Gamma^\hmu \lambda)_\hbeta \cdd_\hmu \nn \\
    &=& (1+ \lambda^\halpha D_\halpha) A_\hbeta
        - 2(\Gamma^\hmu\lambda)_\hbeta A_\hmu.
\eea
The operator ${\bf D}$ acts on a homogeneous polynomial in $\l$
by multiplication by the degree of homogeneity; it does not
change its degree. Therefore, the relations (\ref{drecursion}) are
indeed recursive in powers of $\l$.  They enable us to reconstruct the
superfields
$A_{\hm}$, $A_{\ha}$, and $\Psi^{\ha}$ in their entirety from the
lowest-order data
\be
  A_\hmu = a_\hmu + O(\lambda),\qquad A_\halpha = O(\lambda), \qquad
\Psi^\halpha = \psi^\halpha + O(\lambda).
\label{initialcond}
\ee
The result is
\bea
   A_\hmu & = &  a_\hmu - \lambda \Gamma_\hmu \psi
   - {1\over 4} (\lambda \Gamma_\hmu \Gamma^{\hnu\hat{\rho}}\lambda)
      (f_{\nu\rho} + {2 \over 3} \Gamma_{[\hnu,} \cdd_{\hat{\rho}]} \psi)
+
   \cdots,  \nn \\
 A_\halpha &=& (\Gamma^\hmu \lambda)_\halpha
\left[ a_\hmu
-{2 \over 3} \lambda \Gamma_\hmu \psi
  - {1\over 8}(\lambda \Gamma_\hmu \Gamma^{\hnu\hat{\rho}}
  \lambda) f_{\hnu\hat{\rho}} + \cdots\right], \nn \\
\Psi^\halpha &=&
 \psi^\halpha + {1 \over 2} (\Gamma^{\hmu\hnu} \lambda)^\halpha
 ( f_{\hmu\hnu} - \lambda \Gamma_{[ \hmu,}
  \cdd_{\hnu]} \psi ) +\cdots .
\label{eliminated}
\eea
The superfields in (\ref{eliminated}) are written exclusively in terms
of the physical gauge field $a_{\m}$, its field strength $f_{\m \n}$,
and the gaugino $\psi$; all auxiliary fields have been eliminated.
Moreover, the gauge-fixing condition (\ref{dgauge}) is automatically
satisfied.  This is because the lowest-order data and the first
equation in (\ref{drecursion}) imply that $A_\halpha =
(\lambda \Gamma_\hmu)_\halpha V^\hmu$, for some vector $V^\hmu$.
The condition $\l^{\ha} A_\halpha = 0 $ then follows from
the identity  $\lambda \Gamma^\hmu \lambda = 0$, which is a
consequence of the symmetry of the Dirac matrices,
$\Gamma^\hmu_{\halpha\hbeta}  = \Gamma^\hmu_{\hbeta\halpha}$.

We have not yet exhausted all the relations
that follow from the constraint $F_{\halpha\hbeta}=0$.
Our next task is to show that, if we substitute the solution
(\ref{eliminated}) of the ${\bf D}$-recursion relations into the
constraint,
we obtain the equations of motion for $a_\hmu$ and $\psi$, and nothing
else.
It is possible to calculate $F_{\ha \hb}$ from (\ref{eliminated}) by
brute force.  To the first few orders in $\l$, we find
\be
 F_{\halpha\hbeta} =
(\Gamma_\hmu \lambda)_\halpha (\Gamma_\hnu \lambda)_\hbeta
\left[ -{2\over 15}    \lambda \Gamma^{\hmu\hnu} \dsl \psi
 -{1 \over 18} (\lambda\Gamma^{\hmu\hnu} \Gamma^{\hat{\rho}}\lambda)
     \left(\nabla^{\hat{\sigma}} f_{\hat{\sigma}\hat{\rho}}
  + {1 \over 2} \Gamma_{\hat{\rho}\, \hat{\gamma}\hat{\delta}} \{
 \psi^{\hat{\gamma}}, \psi^{\hat{\delta}} \}\right) + \cdots \right].
\label{eomexpansion}
\ee
This expansion expresses $F_{\ha \hb}$ in terms of quantities that are
set to zero by the \SYM equations of motion (\ref{ymeom}).
Its explicit form will be used in the next subsection
to evaluate the Schwinger-Dyson equation.

Calculating even the first terms of the $F_{\ha \hb}$ expansion in this
manner calls for algebraic heroics.  A more workable method relies on
a set of ${\bf D}$-recursions for the relations
(\ref{impliedbyconstraint}).  The Bianchi identities and the Dirac
matrix properties listed in Appendix A.3 imply
\bea
   (2+{\bf D}) F_{\halpha\hbeta} &=&
  2 (\Gamma^\hmu \lambda)_{\halpha}
 \left( F_{\hmu\hbeta} - (\Gamma_\hmu \Psi)_\hbeta \right)
   + (\halpha \leftrightarrow \hbeta), \nn \\
 (1 + {\bf D}) \left( F_{\hmu\halpha} - (\Gamma_\hmu \Psi)_\halpha \right)
 &=& (\Gamma_\hmu \lambda)_\hbeta \left( \cdd_\halpha \Psi^\hbeta
  - {1 \over 2} \Gamma^{\hmu\hnu}_{\halpha}{}^{\hbeta} F_{\hmu\hnu}
\right),
\nn \\
 (1 + {\bf D}) \left( \cdd_\halpha \Psi^\hbeta
  - {1 \over 2} \Gamma^{\hmu\hnu}_{\halpha}{}^{\hbeta} F_{\hmu\hnu}
\right)
&=&  (\Gamma^{\hmu\hnu} \lambda)^\hbeta
 \cdd_\hmu \left(F_{\hnu\halpha} - (\Gamma_{\hnu}
 \Psi)_\halpha\right) + (\Gamma^\hmu \lambda)_\halpha (\Gamma_\hmu \Dsl
\, \Psi)^\hbeta  \nn \\
&& \qquad \qquad \qquad - \d_{\ha}^{\hb} (\l \Dsl \, \Psi) - \l^{\hb} (\Dsl
\, \Psi)_{\ha}.
\label{constrec1}
\eea
This chain of equations relates $F_{\ha \hb}$ and $F_{\hm \ha}$
to the equation of motion
for the superfield $\Psi$.  To express $F_{\ha \hb}$
and $F_{\hm \ha}$ in
terms of the component field equations of motion, we use the recursion
relations
\bea
  {\bf D} (\Dsl~ \Psi)_\halpha &=&
  (\Gamma^\hmu \lambda)_\halpha ( \cdd^\hnu F_{\hnu\hmu}
  +{1\over 2} \Gamma_{\hmu\hbeta\hat{\gamma}}
  \{ \Psi^\hbeta, \Psi^{\hat{\gamma}} \}), \nn \\
  {\bf D} ( \cdd^\hnu F_{\hnu\hmu}
  +{1\over 2} \Gamma_{\hmu\halpha\hbeta}
  \{ \Psi^\halpha, \Psi^{\hbeta} \})
&=& \lambda \Gamma_{\hmu\hnu} \cdd^\hnu \Dsl~ \Psi ,
\label{constrec2}
\eea
which are also derived from the Bianchi identities and the Dirac
algebra.  Starting with the initial condition (\ref{initialcond}), we
can solve (\ref{constrec1}) and (\ref{constrec2}) iteratively in
powers of $\l$, to express $F_{\halpha\hbeta}$ and $F_{\hm \ha}$
in terms of $a_\hmu$, $\psi$ and their derivatives.   
This procedure generates the
$\lambda$-expansion (\ref{eomexpansion}).  Apart from being
calculationally tractable, it provides a general
proof that the constraint $F_{\halpha\hbeta}=0$ implies
the \SYM \eom and nothing more.

\subsection{The Schwinger-Dyson Equation}

Given our algorithm for expressing $W$ in terms of the physical
fields $a_\hmu$ and $\psi$, we can apply the
Schwinger-Dyson equation to compute
$\delta_\kappa \langle W \rangle$. We wish to evaluate
\be
   \delta_\kappa \langle W \rangle = - \tr P \oint d\tau
  \left\langle \left( p^\hmu  \delta_\kappa \lambda^\halpha
F_{\hmu\halpha}
   + \delta_\kappa \lambda^\halpha \dot \lambda^\hbeta F_{\halpha\hbeta}
\right) W_0 \right\rangle ,
\label{wilsonvevvariation}
\ee
where $W_0 = \ft 1 N \exp \left( \oint d\t \, (A_{\m}
p^{\m} + A_{\a} \dot \l^{\a}) \right)$.
The recursion relations
(\ref{constrec1}) and (\ref{constrec2}) give
$F_{\hmu\halpha}$ and $F_{\halpha\hbeta}$ in terms of
\be
\dsl \psi = -g^2 {\delta S_{SYM} \over \delta \psi}
\ee
and
\be
 \nabla^{\hat{\sigma}} f_{\hat{\sigma}\hat{\rho}}
  + \ft 1 2 \Gamma_{\hat{\rho} \, \hat{\gamma}\hat{\delta}} \{
 \psi^{\hat{\gamma}}, \psi^{\hat{\delta}} \} 
  = g^2 {\delta S_{SYM} \over \delta a_{\hat \rho}},
\label{actionvariation}
\ee
where $S_{SYM}$ is the action for the ${\cal N}=4$ \SYM theory,
and $g$ is the Yang-Mills coupling constant. An integration by parts in the
functional integral transfers the functional
derivatives $\delta/\delta \psi$ and $\delta /\delta a_\hmu$ onto
$W_0$.  Substituting the expansions (\ref{eliminated}) of the gauge
superfields into the definition of $W_0$ then enables us to write the
functional derivatives of $W_0$ in terms of
the physical gauge field and gaugino.  The entire procedure may be
carried out to any desired order in $\l$.

Let us fill in some of the details of the calculation at lowest
nontrivial order in $\lambda$.   At this order, the $F_{\ha \hb}$ term in
(\ref{wilsonvevvariation}) can be neglected, and the expansion of
$F_{\hm \ha}$ can be truncated at quadratic order in $\l$.  From
(\ref{bianchi}) and (\ref{eomexpansion}) we calculate
\bea
F_{\hm \ha} &=& \ft 1 {300} \left( (\l \G_{\hl} \G_{\hm} \G_{\hr} \l)
(\G^{\hl \hr} \dsl \psi)_{\ha} - 4 (\l \G_{\hl} \G_{\hm} \G_{\hr})_{\ha}
(\l \G^{\hl \hr} \dsl \psi) \right. \nn \\
&& \left. \qquad + 2 \, \tr (\G_{\hm} \G_{\hl}) (\l
\G_{\hr})_{\ha} (\l \G^{\hl \hr} \dsl \psi) - 2 (\l \G_{\hr})_{\ha}
(\l \G_{\hl} \G_{\hm} \G^{\hl \hr} \dsl \psi) \right).
\eea
Substituting this expression into
(\ref{wilsonvevvariation}) and applying Dirac matrix identities, we find
\bea
 \delta_\kappa \langle W \rangle &=&
-{1\over 300} \oint d\t \, \left\langle
\left( (\lambda \Gamma_\hmu\psl \Gamma_\hnu \lambda)
   (\kappa\psl \Gamma^{\hmu\hnu} \dsl~\psi) \right. \right. \nn\\
&&~~~~~~ - 4 (\kappa \psl \Gamma_\hmu \psl \Gamma_\hnu \lambda)
    (\lambda \Gamma^{\hmu\hnu} \dsl~\psi) +
64 (\kappa \Gamma_\hmu \lambda)
     (\lambda [\psl, \Gamma^\hnu] \dsl~\psi)  \nn\\
&&~~~~~~ \left. \left.  - 2 (\kappa \psl \Gamma_{\hmu} \lambda)
     (\lambda \Gamma_\hnu \psl \Gamma^{\hnu\hmu}\dsl~\psi)
  \right)
   W_0 \right\rangle.
\label{variationleading}
\eea
Each $\dsl\psi(x(\t))$ is then replaced by
the functional variation
\be
\langle \dsl\psi(x(\t)) W_0 \rangle=
 g^2 {\delta \langle W_0 \rangle \over \delta \psi^\halpha(x(\t))}
 = g^2 \oint d\t' \, \delta(x(\t)-x(\t')) (\psl \lambda)^\halpha(\t')
\langle W_0 \rangle.
\label{schwingerdyson}
\ee
Each of the resulting terms on the right-hand side of
(\ref{variationleading})
is zero. For example,
\bea
(\k \psl \Gamma_\hmu \l) (\l \Gamma_\hnu \psl \Gamma^{\hnu\hmu} \psl \l)
&=& (\k \psl \Gamma_\hmu \l) (\l
\Gamma_\hnu \psl \Gamma^\hnu \Gamma^\hmu \psl \l)
- (\k \psl \Gamma_\hmu \l) (\l \Gamma^\hmu \psl \psl \l) \nn
\\ &=& -8  (\k \psl \Gamma_\hmu \l) (\l \psl \Gamma^\hmu \psl \l)
\nn \\ &=& 0,
\label{vanishing}
\eea
by $p^2=0$ and $\lambda \Gamma^\hmu \lambda = 0$.
We conclude that, if the loop $x(\t)$ is smooth
and has  no nontrivial self-intersections
({\it i.e.}, if $x(\t)=x(\t')$ implies $\t=\t'$),
then the  vacuum expectation value of the Wilson loop
is $\kappa$-invariant,
\be
  \delta_\kappa \langle W \rangle = 0.
\ee

The situation is different when the loop has a self-intersection
point.  In this case, (\ref{variationleading}) yields
\be
  \delta_\kappa \langle W \rangle
 = {{g^2} N\over 6} \oint d\t_1 \oint d\t_2 \, \delta(x(\t_1)-x(\t_2))
   p^\hmu(\t_1) p^\hnu(\t_2) (\lambda \Gamma_{\hmu\hnu\hat{\rho}}
  \lambda)(\kappa \psl(\t_1) \Gamma^{\hat{\rho}} \lambda)
  \langle W_1 W_2 \rangle ,
\label{intersection}
\ee
at leading nontrivial order in the $\lambda$-expansion,
where $W_1$ and $W_2$ are operators associated
to the loops obtained by recombining the original loop
at the intersection.

The integrand is nonzero unless $p(\t_1)$ is parallel to $p(\t_2)$.
Thus the $\kappa$-invariance of the Wilson loop
is broken at self-intersection points. 

\newpage

\section{Matching the Wilson Loop to the String Worldsheet}
\setcounter{equation}{0}
 
 The purpose of this
section is to show how the $\k$-invariance of Wilson loops can be
understood, via the \adscft correspondence, as following from
$\k$-invariance of string theory in $AdS_5 \times S^5$ superspace
\cite{Tseytlin}. 

\subsection{The Wilson Loop as a Boundary Condition}

According to the $AdS/CFT$ correspondence, 
the expectation value of a
Wilson loop operator in strongly coupled four-dimensional ${\cal N} = 4$
\SYM theory can be calculated from the 
worldsheet of type IIB string theory on $AdS_5
\times S^5$ \cite{malda2,reyyee}.  
The \SYM theory is thought of as living on the boundary of
$AdS_5$, and the worldsheet is characterized by the requirement that
it end on the Wilson loop.
To make sense of this conjecture even for bosonic Wilson loops,
it is necessary to think carefully about the boundary conditions the
Wilson loop imposes on the string worldsheet ending on it. This subsection
is
devoted to reviewing these considerations and extending them to the
supersymmetric case.

The supersymmetric Wilson loop is defined along a contour $z^M(\tau)$ in
the
conformal superspace $C_{(10|16)}$ with worldline parameter
$\tau$; that is,
\be
z^M(\tau)=(x^\mu(\tau),y^m(\tau), \lambda^a_\a(\tau))\, .
\ee
According to the conjecture, the loop is the boundary $\s = 0$ of a
string worldsheet
embedded in the $AdS_5 \times S^5$ superspace $C_{(10|32)}$,
with worldsheet parameters $(\tau, \s)$ and embedding coordinates
\be
Z^{\bf M}(\tau,\s) = (X^\mu(\tau,\sigma),Y^m(\tau,\s),
\theta^a_\a(\tau,\s),
\vartheta^a_\a(\tau,\s))\, .
\ee
The boundary conditions imposed by the Wilson loop on the string
worldsheet must therefore translate into relations among the $z^M(\tau)$
and $ Z^{\bf M}(\tau,\s)$ in the limit $\s\rightarrow 0$.

For the bosonic variables, the analysis proceeds exactly as in
\cite{dgo}, where purely bosonic Wilson loops were considered.   It is
natural to impose Dirichlet conditions on $X^\mu$,
\be
\label{xbc} X^\mu(\tau,\s=0) = x^\mu(\tau)\, ,
\ee
identifying the $X^\m$ at the boundary of the worldsheet with the
coordinates in the ${\cal N} = 4$ gauge theory.
The relation between the $y^m$ and $Y^m$ coordinates is more subtle.
It was argued in \cite{dgo}
that the appropriate boundary conditions on the $Y^m$ are
Neumann,
\be
\label{ybc}
P^m_{\t}(\t, \s = 0) = \dot{y}^m(\t) \, ,
\ee
where we have introduced the conjugate momentum
\be
P_{\hm}^i = \frac{\delta {\cal L}}{\delta (\partial_i X^{\hm})}~~~
(i = \tau, \sigma),
\ee
and ${\cal L}$ is the string Lagrangian.  The momentum satisfies
\be
P^{\un \hm}_i = J_i^j E_j^{\un \hm},
\ee
where
\be
J_i^j = \frac{g_{ik} \e^{kj}}{\sqrt g} \label{compstr}
\ee
is the worldsheet complex structure, written in terms of the
worldsheet metric $g$ and the antisymmetric tensor density $\e^{ij}$
($\e^{\t \s} = +1$).

Now to the fermionic variables. Our proposed boundary conditions for
the fermions stem from the result, proven in section 2,
that at the boundary $Y = 0$ of $AdS_5 \times S^5$ superspace, the
$\vt$  coordinates decouple from the superisometry variations
of the remaining coordinates.   Moreover, the variations of the
$\theta$'s  reduce
at the boundary to the variations under superconformal transformations
of the $\lambda$'s of the $(10|16)$-dimensional conformal superspace.
This strongly suggests the boundary condition\footnote{This equation
is not precise: as discussed in Appendix A.2, the spinor $\th_{\a}$ is
chiral under the Lorentz group $SO(1,3)$ of the boundary, whereas
$\l_{\a}$, as defined in section 3, is Majorana. This distinction will
not matter for us until we compare $\k$-symmetries in section 4.2.}
\be
\label{thetabc}
\theta^a_\a(\tau,\s=0) = \lambda^a_\a(\tau) \, .
\ee

What conditions must we impose at $Y=0$ on the $\vt$'s?  We claim
that no further conditions are necessary.  This may be understood from
general features of the equations of motion for the string worldsheet
in $AdS_5 \times S^5$ superspace \cite{Tseytlin}.  The equations
for the bosonic coordinates are second-order; those for the fermions
are first-order; and the entire system is of elliptic type.  A
second-order elliptic system in a given region is completely specified
by giving one piece of boundary data ({\it i.e.}, a Dirichlet or Neumann
condition) per variable, and indeed we found that the six Neumann
conditions
(\ref{ybc}) and four Dirichlet conditions (\ref{xbc}) are just sufficient
to determine the ten bosonic worldsheet coordinates $X^\m(\t, \s)$ and
$Y^m(\t, \s)$.  On the other hand, since the 32 fermionic equations are
first-order, and two first-order equations are generically equivalent
to one second-order equation, we expect that only 16 boundary
conditions are required to determine all 32 fermionic coordinates
$\theta$ and $\vt$.  Having already supplied 16 boundary
conditions in (\ref{thetabc}), we need provide no further boundary
data: $\theta(\t, \s)$ and $\vt(\t, \s)$ are uniquely fixed by the
string equations of motion and the boundary values of $\theta(\t,
\s = 0)$.  This point of view is consistent with the boundary
decoupling of $\vt$ discovered in section 2.

Of course, this counting cannot be the whole story, since we have ignored
$\k$-symmetry.  Though the string worldsheet propagates in a superspace of
32
fermionic dimensions, 16 of these degrees of freedom are gauge
artifacts, removable by fixing $\k$-symmetry.  As we shall show in the
next subsection, the 16-dimensional $\k$-symmetry decomposes in such a
way that eight independent symmetries act on the $\theta$ coordinates and
eight on the
$\vartheta$'s.  Thus, if we fix $\k$-symmetry entirely, we are left
with 16 independent fermionic degrees of freedom: eight $\theta$'s and eight
$\vartheta$'s.  By the argument of the last paragraph, these are
completely determined everywhere on the worldsheet by the values at
the boundary $\s = 0$ of the eight ``unfixed'' $\theta$'s.  This is
consistent with our expectations from \SYM theory.  The worldline of a
massless particle in ten dimensions coupled to a background gauge field
is apparently $(1|16)$-dimensional, but this system, too, admits a
$\k$-symmetry \cite{Rocek}, which cuts the number of effective
fermionic degrees of freedom to eight.  At the boundary of the string
worldsheet, the eight ``unfixed'' $\theta$'s may be identified with these
eight $\lambda$'s unfixed by $\k$-symmetry of the Wilson loop worldline.

The arguments we have just given, together with arguments made in
\cite{dgo} for the bosonic variables, demonstrate that the boundary
conditions (\ref{xbc}), (\ref{ybc}), and (\ref{thetabc}) suffice to
determine a minimal supersurface in 
$AdS_5 \times S^5$ superspace. However, the prescription of
\cite{malda2,reyyee,dgo} demands in addition 
that the boundary $\s = 0$ of the worldsheet end on
the boundary $Y^m = 0$ of $AdS_5$.  This is not
guaranteed from the boundary conditions alone.  We now show that a
necessary condition for the worldsheet to terminate on
the boundary of $AdS_5$ is the condition $p^2 = 0$ on the loop
variable.  More
specifically, we show that the
Virasoro constraint of the string, evaluated at the boundary of the
worldsheet, is equivalent to the condition $p^2 = 0$, up to terms that
vanish at $Y=0$.

It is convenient to write the Virasoro constraints
as
\bea
0 &=& 2 E_i^{\underline{\hat \mu}} E_{j \underline{\hat \mu}} -  g_{ij} g^{kl} 
E_k^{\underline{\hat \mu}} E_{l
\underline{\hat \mu}} \nn \\
&=& P_i \cdot P_j + E_i \cdot E_j, ~~~~~~~~~~~~~(i,j=\tau,\sigma) ,
\eea
where we have used the identity
\be
{1\over g} \e^{ij} \e^{kl} = g^{ik}g^{jl} - g^{il}g^{jk}.
\ee
We will show in the following paragraphs that 
\be
\label{whatweneed}
{P^{\un \mu}_\tau P_{\tau \un \mu} \over E^{\un \mu}_\tau E_{\tau \un \mu}} 
= 0 
\qquad \hbox{and} \qquad
{E^{\un m}_\t E_{\tau \un m} \over P^{\un m}_\tau P_{\tau \un m}} = 0, 
\ee
in the limit $Y \to 0$.
Granting this, the Virasoro constraints reduce to the
single equation
\be \label{boundaryvir}
P_\t^{\un m} P_{\t \un m} + E_\t^{\un \m} E_{\t \un \m} = 0
\ee
at the boundary. This equation is equivalent to the condition $p^2=0$ on 
the Wilson loop. 
To see this, we use the form of the vielbeins given in section 2.4
and the boundary conditions on the string worldsheet
to rewrite (\ref{boundaryvir}) as
\be
\label{Virasoro}
P_\t^{\un m} P_{\t \un m} +
E^{\un \m}_{\t} E_{\t \un \m}  = Y^{-2} \left[( \dot{x}^{\un \m} +
\ft 1 2 (\dot{\bar{\l}} \g^{\un \m} \l) - \ft 1 2 (\bar{\l}
\g^{\un \m} \dot{\l}))^2 + \dot{y}^2\right] + O(1)=0.
\ee 
We then compare this with
\bea
\label{PRedef}
p^{\un \m} &=& \dot{x}^{\un \mu} + \ft 1 2 \dot{\bar{\lambda}}  \G^{\un \m} \l  - \ft 1
2 \bar{\lambda}  \G^{\un \m} \dot{\l}, \nn \\
p^{\un m} &=& \dot{y}^{\un m},
\eea
where we have taken into account the redefinition of $\dot y$ discussed at
the beginning of section 3. 
Working in the ``5+5'' basis for the
$\G$-matrices described in Appendix A.1,  we find
\be
p^2 = (\dot{x}^{\un \m} + \ft 1 2 \dot{\bar{\l}} \gamma^{\un \m} \l - \ft
1 2 \bar{\l} \gamma^{\un \m} \dot{\l})^2 + \dot{y}^2.
\ee
Upon multiplication by $\ft 1 {Y^2}$, this is identical to the leading
term in
(\ref{Virasoro}). Thus the Virasoro condition of the string,
restricted to the boundary, is equivalent to the condition $p^2 = 0$.

We now prove (\ref{whatweneed}), assuming smooth worldsheet
boundary conditions.  Let us first describe the idea of the proof.  We
are interested in understanding how the worldsheet behaves as it
approaches the boundary at $Y^m = 0$.  If the worldsheet boundary is
constrained to be bosonic and straight,
\bea
\label{bosonic&straight}
\l(\t, \s = 0) &=& 0, \nn \\
x^{\hm}(\t, \s = 0) &=& v^{\hm} \t,
\eea
then the worldsheet itself must be approximately flat near the
boundary and perpendicular to it: fluctuations in the worldsheet 
geometry are energetically
costly, on account of the factor $Y^{-2}$ in the $AdS_5$ metric.  We
claim that, no matter what the worldsheet boundary condition actually
is, the worldsheet near any given point $\t = \t_0$ on the
boundary can be locally approximated by a worldsheet obeying
(\ref{bosonic&straight}).

Let us now make this precise. Suppose $(X^{\m}, Y^m, \th)$
is a solution of the string \eom with general boundary conditions
\bea
X^{\m}(\t, \s = 0) &=& v^{\m} \t + \ft 1 2 a^{\m} \t^2 + \cdots, \nn
\\
P^{m}_{\t}(\t, \s = 0) &=& v^m +  a^m \tau + \cdots, \nn \\
\th(\t, \s = 0) &=& \l_1 \t + \l_2 \t^2 + \cdots, \nn \\
Y^m(\t, \s = 0) &=& 0.
\eea
We have set $\tau_0=0$ for simplicity. In addition, 
by suitably positioning the origin of the boundary coordinates, we have
assumed $x^{\m} = \lambda = 0$ at $\t = 0$.
Since the string worldsheet
action in $AdS_5$ is scale-invariant, $(\widetilde
X^{\m}, \widetilde Y, \widetilde \th) = (\Omega X^{\m}, \Omega Y, \Omega^{1/2}
\th)$ is also a solution, with the boundary conditions
\bea
\widetilde X^{\m}(\t, \s = 0) &=& \Omega v^{\m} \t + \ft 1 2 \Omega a^{\m}
\t^2 + \cdots, \nn
\\
\widetilde P^{m}_{\t}
(\t, \s = 0) &=& \Omega v^m + \Omega a^m\tau + \cdots, \nn \\
\widetilde \th(\t, \s = 0) &=& \Omega^{1/2} \l_1 \t + \Omega^{1/2} \l_2 \t^2 +
\cdots, \nn \\
\widetilde Y^m(\t, \s = 0) &=& 0.
\eea
The action moreover is invariant under worldsheet reparametrizations.
The reparametrization $\t \to \hat{\t} = \Omega\tau$ 
transforms the boundary conditions to
\bea
\widetilde X^{\m}(\hat \t, \s = 0) &=& v^{\m} \hat\t + \ft 1 2 a^{\m} 
{\hat{\t}^2 \over \Omega} +
\cdots, \nn
\\
\widetilde P^{m}_{\hat{\t}}(\hat \t, \s = 0) &=& v^m + a^m {\hat \t \over \Omega}
+\cdots, \nn \\
\widetilde \th(\hat\t, \s = 0) &=& \Omega^{-1/2} \l_1 \hat\t 
+ \Omega^{-3/2} \l_2 \hat\t^2 +
\cdots, \nn \\
\widetilde Y^m(\t, \s = 0) &=& 0.
\label{foureighteen}
\eea

The expressions in (\ref{whatweneed}) are invariant under both the
rescaling of $(X^{\m}, Y^m, \th)$ and the reparametrization of $\t$,
for any $\Omega$.  In the limit  $\Omega \to \infty$, the boundary
conditions (\ref{foureighteen}) become those of a 
straight and bosonic worldsheet,
\bea
\label{straightboundary}
\hat X^{\m}(\hat\t, \s = 0) &=& v^{\m} \hat\t, \nn \\
\hat P^{m}_{\t}(\hat\t, \s = 0) &=& v^m, \nn \\
\hat \th(\hat \t, \s = 0) &=& 0, \nn \\
\hat Y^m(\hat\t, \s = 0) &=& 0.
\eea

It is therefore sufficient to establish (\ref{whatweneed}) 
for worldsheets obeying
the simple boundary conditions (\ref{straightboundary}).
First of all, since
$Y^m=0$ at the boundary, clearly $E^{\un m}_\tau \sim \partial_\tau Y^{\un m} 
=0$.  
Furthermore, it was shown in \cite{malda2} that worldsheets obeying
(\ref{straightboundary}) satisfy
\be
X \sim Y^3 \rightarrow 0
\ee
near the boundary. 
Computing $J_i^j$ and $E_i^{\un \m}$ is then
straightforward, and shows that $P_{\t}^{\un \m} = 0$ at the boundary.

We have shown that, for a smooth Wilson loop, the condition
$p^2=0$ is necessary in order for the string worldsheet to end
on the boundary $Y=0$. It is interesting to note that, for
the same reason as found in \cite{dgo}, the agreement between
the $p^2=0$ condition of the loop and the Virasoro constraints
of the string worldsheet fails when the loop has intersections. 
The disagreement between the $p^2=0$ condition and the
Virasoro constraint at these points may be a cause of 
the breakdown (\ref{intersection}) of $\kappa$-invariance.

\subsection{Matching the Kappa Symmetries}

The prescription of the previous subsection implies that
the Wilson loop expectation value $\langle W \rangle$ is obtained
as a functional integral over string worldsheets obeying
the boundary conditions (4.3), (4.4), and (4.8). We may thus
view $\langle W \rangle$ as a wave function of Hartle-Hawking type,
with no boundary other than the Wilson loop itself. Since
the action of \cite{Tseytlin} is invariant under $\kappa$-symmetry,
this wave function must obey a set of constraints, corresponding to
the vanishing of the momenta conjugate to the directions of the
$\kappa$-symmetry. This is a standard statement in any gauge
theory. In Maxwell theory, for example, the momentum conjugate to the 
timelike component $A_0$ of the gauge field vanishes,
and the wave function $\Psi$ of the theory obeys the constraint
$\delta \Psi /\delta A_0 = 0$. We claim that the constraint 
due to the $\kappa$-symmetry of the worldsheet is nothing
but the equation
\be
\label{againkappa}
 \delta_\kappa \langle W \rangle = 0
\ee
obtained in section 3 within gauge theory. 

To show that the constraint from the worldsheet $\kappa$-symmetry
is the same as (\ref{againkappa}), it is sufficient to check that 
the $\k$-symmetries of the string worldsheet reduce at 
the boundary to the worldline $\k$-symmetries of the Wilson 
loop. 
The $\k$-transformations of the string 
propagating in the $AdS_5 \times S^5$ superspace given in \cite{Tseytlin} 
read
\bea
\label{MTkappa}
\delta_\kappa Z^{\bf M} E_{\bf M}^I&=& 2 {E^{\underline
{\hat \mu}}}_i
{\hat \Gamma_{\underline {\hat \m}}} \kappa^{Ii},\nn \\
\delta_\kappa Z^{\bf M} E_{\bf M}^{\underline {\hat \mu}}&=&0\, .
\eea
Here $I = 1,2$  labels the two Majorana-Weyl
fermionic generators of Type IIB supergravity on $\adsxs$. The
corresponding fermionic vielbeins $E^{1,2}$ are related to the fermionic
vielbeins defined in section 2 by
\be
E_{Q/S}=\frac{1}{2\sqrt{2}}\left(1\mp\gamma^5\right)(E^1+i E^2)\, .
\ee 
The $\hat \G$-matrices are the ones defined in
Appendix A.1.  The $\k$-symmetry parameters are packaged
in  two Majorana-Weyl quantities $\k^{1i}$
and  $\k^{2i}$, each of which carries a (hidden) 
spinor index as well as a (visible) worldsheet vector
index $i = (\tau, \sigma)$.  The $\k^{Ii}$ obey the
worldsheet self-duality relations
\bea
\k^1 &\equiv & \kappa^{1 \tau}=J_i^\tau \kappa^{1 i}, \nn \\
\k^2 &\equiv & \kappa^{2 \tau}=-J_i^\tau \kappa^{2 i}  \, ,
\eea
where $J_i^j$ is the complex structure defined in (\ref{compstr}). 
We evaluate the $\k$-variations in the boundary limit $Y \to 0$. For 
simplicity, we 
consider only the case of constant $\dot Y^i / |\dot Y|$; that is, we take
the worldsheet to be located at a fixed point on $S^5$.

As we have remarked, in the
limit $Y \to 0$, the vielbein components $P^{\un \m}_{\sigma} = 
J_\sigma^i E^{\un \m}_i$ 
and $E^{\un Y}_{\tau}$ decouple.  The restriction to the $AdS_5$
directions entitles us to replace $(\G^\m, \G^Y) \to (\g^\m, \g^5)$, as
explained in Appendix A.1.  Expanding the first equation in
(\ref{MTkappa}) subject to these assumptions gives
\bea
\delta_\kappa Z^{\bf M} E_{\bf M}^1&=& 2 
(E^{\un \m}_{\tau} \g_{\un \m} - P^{\un Y}_{\tau} \g_5) 
\k^1, \nn \\
\delta_\kappa Z^{\bf M} E_{\bf M}^2&=& 2 
(E^{\un \m}_{\tau} \g_{\un \m} + P^{\un Y}_{\tau} \g_5) \k^2.
\eea
Let us define $\k \equiv \ft 1 {\sqrt 2}(\k^1 + i \k^2)$, $\widetilde \k
\equiv \ft 1 {\sqrt 2}(\k^1 - i \k^2$), $\k_\pm \equiv \ft {1 \pm
\g^5} 2 \k$,  and $\widetilde \k_\pm
\equiv \ft{1 \pm \g^5} 2  \widetilde \k$.  Then
\bea
\d Z^{\bf M} E^{ a}_{Q\bf M }&=& 
2 \left(E^{\un \m}_{\tau} (\g_{\un \m} \k_+)^a +
P^{\un Y}_\tau  \widetilde{\k}_-^a \right), \nn \\
\d Z^{\bf M} E^{a}_{S \bf M}&=& 
2 \left(E^{\un \m}_{\tau} (\g_{\un \m} \k_-)^a -
P^{\un Y}_\tau  \widetilde{\k}_+^a \right).
\eea
Here and in many subsequent formulas, the $SO(1,4)$ spinor index has
been suppressed for readability.

It will be useful for what follows to work out the properties of the
various $\k$'s under complex conjugation.  The $\k^I$ are
Majorana spinors in ten dimensions; therefore, as discussed in Appendix
A.2, $(\k^I)^* = (B \otimes B') \k^I$.  It follows that $\k^* =
(B \otimes B') \widetilde \k$; also,
\be
\widetilde{\k}_- = \ft {1 - \g^5} 2 (B^\dag \otimes B'^\dag) \k^* =  (B^\dag
\otimes B'^\dag) \ft {1 + \g^5} 2 \k^* =   (B^\dag \otimes
B'^\dag) (\k_+)^*,
\ee
where the second equality is true because $\gamma^5$ and $B$ anticommute, and 
the last step follows because $\g^5$ is real in our chosen
representation. 
Similarly, $(\widetilde \k)_+ = (B^\dag \otimes
B'^\dag)(\k_-)^*$.
The variations then become
\bea
\label{dkappatheta}
\delta_\kappa Z^{\bf M} (E_{Q}^a)_{\bf M} &=&
2\left(E^{\un \m}_{\tau}
(\g_{\un \m} \k_+)^a  + P^{\un Y}_\tau ((B^\dag \otimes B'^\dag) (\k_+)^*)^a
\right), \\
\delta_\kappa Z^{\bf M} (E_{S}^a)_{\bf M}&=& 2
\left(E^{\un \m}_{\tau}
(\g_{\un \m} \k_-)^a  - P^{\un Y}_\tau ((B^\dag \otimes B'^\dag) (\k_-)^*)^a
\right).
\eea
From  (\ref{BoundVielbein}) and (\ref{MTkappa}) it follows that
\be
\d_\kappa Z^{\bf M}(E_{Q}^a)_{\bf M} \sim\d_\kappa Z^{\bf M} e^{Q a}_{\bf M}=
Y^{-\frac{1}{2}} \delta_\kappa \theta^b u(\phi)_b{}^a\, ,
\label{kappaDecompo}
\ee
where $\sim$ reminds us that we neglect terms which are subleading
in the boundary limit.
With this form it becomes clear that
it is  $\k_+$ which acts on the coordinate $\theta$.

On the other hand, by substituting the bosonic part of the
boundary vielbein (\ref{BoundVielbein}) into (\ref{dkappatheta}), we arrive
at
\be
\label{AdSkappa}
\d Z^{\bf M} E^{Q a}_{\bf M} \sim  \ft 2 {Y} \left[ (\dot x^{\un \m} +
\ft 1 2 \dot{\bar{\theta}} \g^{\un \m} \theta
- \ft 1 2 \bar{\theta} \g^{\un \m} \dot{\theta}) (\g_{\un \m} \k_+)^a
+ P_\tau^{\un Y} \left((B^\dag \otimes B'^\dag) \k_+^*\right)^a \right].
\ee
Combining (\ref{kappaDecompo}), (\ref{AdSkappa}) and the boundary
conditions
\be
\theta|_{\sigma=0}=\lambda, \qquad P_\tau^{Y}=\dot y
\ee
yields the result
\be
\label{CFTkappa}
\d \l^a = (\dot x^{\un \m} +
\ft 1 2 \dot{\bar{\l}} \g^{\un \m} \l
- \ft 1 2 \bar{\l} \g^{\un \m} \dot{\l}) (\g_{\un \m} \k_{SYM})^a  + \dot y 
\left((B^\dag
\otimes
B'^\dag) \k_{SYM}^*\right)^a\, ,
\ee
with
\be
\k_{SYM}^a = \ft 2 {Y^{1/2}} (\k_+)^b (u^{-1})_b{}^a\, .
\ee
By appendices A.1 and A.2, (4.34) is precisely the
$\k$-variation (3.10) of the fermionic gauge theory coordinates, written
in dimensionally reduced form\footnote{Strictly speaking, (\ref{CFTkappa})
gives only the transformation of the chiral component of the Majorana
spinor $\l$. The transformation of the anti-chiral component
follows from the Majorana condition.}.
The proper variations of the bosonic coordinates in (\ref{firstkappa})
follow from the second equation in (\ref{MTkappa}) and the bosonic
boundary vielbein (\ref{BoundVielbein}). In particular, with $E^{\un Y}$
from (\ref{BoundVielbein}) we find
\be
\d_{\kappa} Y=0\, ,
\ee 
which says that a $\kappa$-variation does not remove the
endpoints of the string worldsheet from the boundary of the
$AdS$ space.

To conclude, we have succeeded in deriving the $\kappa$-symmetry 
of the Wilson loop as the restriction of the stringy $\kappa$-symmetry
to the boundary of $AdS_5$.

\newpage
\section{Discussion}

In this paper, we studied how
holography works in theories formulated in 
superspace. We then applied our results to the
computation of Wilson loop expectation values in ${\cal N}=4$
super-Yang-Mills theory in four dimensions. 
We found that the expectation value of a loop
is $\kappa$-invariant, provided the loop is
smooth and lightlike, and we identified this invariance with
the $\kappa$-invariance of the string worldsheet
action. 

Intriguingly, the field theory computation
shows that $\kappa$-symmetry is broken at intersections
of the loop, as we saw in (\ref{intersection}). 
It would be interesting
to derive the same result from the
point of view of the string worldsheet in $AdS_5 \times S^5$.
The breakdown of $\k$-invariance in the loop may be
related to the failure of the proof of (\ref{whatweneed}) at intersections.

The structure of equation (\ref{intersection}) is
similar to that of the loop equation of Makeenko and Migdal
\cite{migdal}. Classically, $\delta_\kappa W = 0$ is equivalent
to the super-Yang-Mills equations of motion, so we expect that
(\ref{intersection}) carries as much information as the
loop equation. Since $\kappa$-variations have a well-defined
geometric meaning in loop space, $\delta_\kappa W$ does
not suffer from the subtlety that arises in defining the loop differential
operator. 

The relation we have studied between bulk and boundary superspaces
seems closely connected to the
relation between gauged supergravity in $AdS$
and superconformal supergravity on the boundary
\cite{liutseytlin}. It would be interesting to understand
this connection better. 

\section*{Acknowledgments}

We would like to thank Martin Cederwall,
Piet Claus, Djordje Minic, John Schwarz, 
and Dmitri Sorokin for useful discussions.

\medskip

\noindent
This research was supported in part by the Caltech
Discovery Fund. \\
H.O., H.R., and J.T. 
are also supported in part by
NSF grant PHY-95-14797 and DOE grant DE-AC03-76SF00098.
J.R was also in part supported by the DOE grant DE-FG03-92ER40701.

\newpage

\appendix
\noindent

 \section{Dirac Matrices and Spinors in Ten Dimensions}
\setcounter{equation}{0}

\subsection{Dirac Matrices}

In this subsection, we explain the two decompositions of the $SO(1,9)$
Dirac matrices that are used in the body of the paper.

We may write the general $32 \times 32$ Dirac matrix in ten dimensions
in the chiral basis
\be
\label{chiralgamma}
\hat{\G}^{\hm} = \left( \matrix{0 & \G^{\hm} \cr
                                \G^{\hm} & 0 \cr} \right) = \G^{\hm}
\otimes \s^1 ,
\ee
where $\G^{\hm}$ is a $16 \times 16$ block and $\s^1$ is a Pauli
matrix.
This form is appropriate when we are dealing with spinors in
ten dimensions of definite $SO(1,9)$ chirality, such as the
spinor introduced in section 3 to parametrize worldline $\k$-symmetry
of the Wilson loop.  Accordingly, the $\G$-matrices used in that section
are the $16 \times 16$ blocks in (\ref{chiralgamma}).  Further
properties of these matrices are listed in Appendix A.3.

Studying spinors in $AdS_5 \times S^5$ necessitates a different and
more refined decomposition, which accommodates the breaking of Lorentz
symmetry from $SO(1,9)$ to $SO(1,4) \times SO(5)$:
\bea
\label{10to4+1+5}
\hat{\G}^{\m} &=& \g^{\m} \otimes {\bf 1} \otimes \s^1, \nn \\
\hat{\G}^Y &=& \g^5 \otimes {\bf 1} \otimes \s^1, \nn \\
\hat{\G}^{m'} &=& {\bf 1} \otimes \g'^{m'} \otimes \s^2.
\eea
Here the $\g^{\m}$ and $\g^5$ are the $4 \times 4$ Dirac matrices of
$SO(1,4)$ (chosen so that $\g^5$ is real), the $\g'^{m'}$ are
$4 \times 4$ Dirac matrices of $SO(5)$,
${\bf 1}$ is the identity matrix in four dimensions, and $\s^1$ and $\s^2$
are
Pauli matrices.   These are the matrices that appear in the
Metsaev-Tseytlin formulation of $\k$-symmetry reviewed in section 4.2.

\subsection{Spinors}

Typically, in constructing type II Green-Schwarz superstring theory,
we take our fermionic coordinates to be two Majorana-Weyl spinors
$\Theta^I$ ($I = 1,2$) of $SO(1,9)$.  The information contained
in these spinors can be repackaged in a single chiral (but no longer
Majorana) spinor
\be
\Theta = \ft 1 {\sqrt{2}}(\Theta^1 + i \Theta^2).
\ee
The spinor $\Theta$ decomposes under $SO(1,9) \to SO(1,4) \times SO(5)$ as
\be
\Theta^{\ha} \to \Theta^a_{\a},
\ee
where $\ha = 1,\dots, 16$ is a (complex-valued) spinor index of
$SO(1,9)$,  and $\a = 1,\dots,4$ and $a = 1,\dots 4$ are spinor
indices  of $SO(1,4)$ and $SO(5)$, respectively.  The conjugate spinor
is defined by
\be
\bar{\Theta}^{\a}_a = i ((\Theta^a)^\dag \g_0)^{\a}.
\ee
It is often useful
to introduce a notion of chirality with respect to an $SO(1,3)$
subgroup of the $SO(1,4)$.  From the standpoint of $SO(1,4)$, {\it
i.e.}, of physics in $AdS_5$, this chirality is completely
fictitious.  However, the \adscft correspondence distinguishes the 4
coordinates $X^{\m}$ of $AdS_5$ parallel to the boundary, and in the
space of these coordinates, $SO(1,3)$ chirality is a natural concept,
implemented by the matrix $\g^5$.
Accordingly, we define the projected spinors
\bea
\label{ourfermions}
\theta^a_{\a} &=&  Y^{\frac{1}{2}} 
\left({1 - \g^5 \over 2}
\right)_\a^\b \Theta_{\b}^b (u(\phi)^{-1})_b^a \, , \nn \\
\vt^a_{\a} &=&   Y^{-\frac{1}{2}} \left( {1 + \g^5 \over 2}
\right)_\a^\b \Theta_{\b}^b (u(\phi)^{-1})_b^a.
\eea
These are the coordinates we work with in section 2.4. The matrices
$u(\phi)$ are the coset representatives of $SO(6)/SO(5)$. These
coordinates are similar, but not identical to the Killing coordinates
introduced in \cite{NearHorizon}.

We conclude this discussion with some remarks on complex conjugation.
It is possible to define a unitary $32 \times 32$ matrix ${\cal B}$ of
complex conjugation, with the property that
\be
\label{Gstar}
(\hat{\G}^{\hm})^* = {\cal B} \hat{\G}^{\hm} {\cal B}^{-1}.
\ee
The complex conjugate of a Majorana spinor $\z$ in ten dimensions is then
\be
\z^* = {\cal B} \z.
\ee
In the basis (\ref{10to4+1+5}), the matrix of complex conjugation becomes
\be
{\cal B} = B \otimes B' \otimes \s^3,
\ee
where $B$ and $B'$ are the unitary matrices of complex conjugation in
five-dimensional Minkowski and Euclidean spaces, respectively, and $\s^3$
is a Pauli matrix.  We do not need the explicit forms of the matrices $B$
and $B'$, although we will use the relation
\be
\label{Bg5}
(1 + \g^5) B = B(1 - \g^5),
\ee
which follows from (\ref{Gstar}) with $\hm = Y$.

The complex conjugate of a Majorana-Weyl spinor $\z_{\a}^a$ is
given in the basis (\ref{10to4+1+5}) by
\be
\label{Majorana}
(\z_{\a}^a)^* = B_{\a}^{\b} B'_b{}^a \z^b_{\b}.
\ee
If we further decompose $\z$ according to the $SO(1,3)$ chirality
described above,
\bea
(\z_+)_{\a}^a &=& \ft 1 2 (1 + \g^5)_{\a}^{\b} \z_{\b}^a \, ,\nn \\
(\z_-)_{\a}^a &=& \ft 1 2 (1 - \g^5)_{\a}^{\b} \z_{\b}^a \, ,
\eea
then unitarity, (\ref{Bg5}), and the Majorana condition
(\ref{Majorana}) imply  the
relation
\be
\label{zetarelation}
\z_- = (B^\dag \otimes B'^\dag) (\z_+)^*.
\ee

\subsection{Dirac Matrix Identities}

In this subsection, we present a list \cite{hs} of Fierz and other
identities satisfied by the 16-dimensional chiral $\G$-matrices
defined in Appendix A.1.  These identities are
used ubiquitously (if unostentatiously) in deriving the various
results of section 3.

\bea
\G^{\hm \, \ha \hb} = \G^{\hm \, \hb \ha}, \qquad \G^{\hm}_{\ha \hb}
&=&   \G^{\hm}_{\hb \ha} \qquad \qquad \qquad \hbox{(symmetry)} \\
\G^{\hm \, \ha \hb} \G^{\hn}_{\hb \hg} + \G^{\hn \, \ha \hb}
\G^{\hm}_{\hb \hg} &=&  2 g^{\hm \hn} \d_{\hg}^{\ha} \qquad \quad
\hbox{(Clifford algebra)} \\
\G^{\hm}_{\ha \hb} \G_{\hm \, \hg \hd} +   \G^{\hm}_{\ha \hg} \G_{\hm
\, \hb \hd} + \G^{\hm}_{\ha \hd} \G_{\hm \, \hb \hg} &=& 0 \qquad
\qquad \qquad \hbox{(Fierz identity)} \\
\G^{\hm}_{\ha \hb} \G_{\hm}^{\ha \hg} &=& 10 \d_{\hb}^{\hg} \\
\G^{\hm}_{\ha \hb} \G_{\hn}^{\ha \hb}  &=& 16 \d_{\hn}^{\hm} \qquad
\qquad \qquad \quad \hbox{(trace)} \\
\G_{\hm} \G^{\hn} \G^{\hm} &=& -8 \G^{\hn} \\
\G^{\hm \hn}{}_{\hb}^{\ha} &=& \ft 1 2 (\G^{\hm \, \ha \hg} \G^{\hn}_{\hg
\hb} -  \G^{\hn \, \ha \hg} \G^{\hm}_{\hg \hb})
\eea

\newpage
\section{The $SU(2,2|4)$ Algebra}
\setcounter{equation}{0}

The supergroup $SU(2,2|4)$ is generated by: conformal translations
$P_{\m}$; Lorentz transformations $M_{\m \n}$; the dilatation
generator $D$;  $SU(4)$ rotations
$U_i^j$ ($i, j = 1,\dots,4$);
ordinary supersymmetries $Q_{\a}^a$; and special supersymmetries
$S_{\a}^a$. 
The
generators of $SU(4)$ rotations may be written as $U_i^j = 2 
\widetilde{P}_{m'}
(\widetilde{\G}^{m' 6})_i^j + \widetilde{M}_{m' n'} 
(\widetilde{\G}^{m' n'})_i^j$, where the 
$\widetilde{P}_{m'}$
and $\widetilde{M}_{m' n'}$ 
($m',n' = 1, \dots, 5$) are generators of translations
and rotations on $S^5$, and the $\widetilde{\G}$'s are the $4 \times 4$
chiral blocks of the $SO(6)$ Dirac matrices in the chiral basis.
The generators are assigned weights according to their
commutation
relations with the dilatation operator: the $P$'s have weight 1; the
$M$'s,  the $U$'s, and  $D$ itself have weight 0; the $K$'s have weight
-1;
the $Q$'s have weight 1/2; and the $S$'s have weight -1/2.  The full
structure of the algebra is

\begin{eqnarray}
{}[ M_{ m n},  M_{ p q}] &=&
\eta_{m[ p}  M_{ q] n} -  \eta_{
n[ p}  M_{ q] m}\,, \nonumber\\
{}[ P_{ q}, M_{ m  n}]=  \eta _{ q [ m}
 P_{ n]},&& {}[ K_{ q}, M_{m n}]=\eta _{ q [ m}
 K_{ n]} \nn \\
{}[D, P_{m}]= P_{ m},&& {}[D, K_{m}]=-K_{m} \nn \\
{}[ P_{m}, K_{n}]&=&2\left(\eta_{mn} D+2 M_{mn}\right)\nn \\
{}[ M_{
m n}, Q_{\a}{}^i] = -\frac 14
(\gamma_{ m n} Q^i)_{\alpha}\,, & &
{}[ M_{
m n}, S_{\a}{}^i] = -\frac 14
(\gamma_{ m n} S^i)_{\alpha}\,
\nonumber\\
{}[ P_{
m}, S_{\a}{}^i] =
(\gamma_{m} Q^i)_{\alpha}\,, &&
{}[ K_{
m}, Q_{\a}{}^i] =
(\gamma_{m} S^i)_{\alpha}\,
\nonumber\\
{}[D, Q_{\a}{}^i] =\frac{1}{2}
Q^i_{\alpha}\,, &&
{}[D, S_{\a}{}^i] =-\frac{1}{2}
S^i_{\alpha}\,,
\nonumber\\
 \{
{Q}_{\alpha}{}^i, \bar {Q}_j{}^{\beta} \}=\d_j{}^i (\gamma^m)_\a{}^b P_m\,
,
&&
\{
{S}_{\alpha}{}^i, \bar {S}_j{}^{\beta} \}=\d_j{}^i (\gamma^m)_\a{}^b K_m\,
, \nn \\
\{
{Q}_{\alpha}{}^i, \bar {S}_j{}^{\beta} \} &=&
\delta_j{}^i D +\delta_j{}^i (\gamma^{ mn})_{\alpha}{}^{\beta}M_{mn}
-2
\delta_{\alpha}{}^{\beta} U_j{}^i\,,\nonumber\\
{}[U_i{}^j, {Q}_{\alpha}^k] &=&  \delta_i{}^k
{Q}_{\alpha}{}^j - \frac 1{4} \delta_i{}^j {Q}_{\alpha}{}^k\,, \nn \\
{}[U_i{}^j, {S}_{\alpha}^k] &=&  \delta_i{}^k
{S}_{\alpha}{}^j - \frac 1{4}
\delta_i{}^j {S}_{\alpha}{}^k\,, \nonumber\\ {}[U_i{}^j, U_k{}^l] &=&
\delta_i{}^l U_k{}^j -  \delta_k{}^j U_i{}^l\,,
\label{SU224SC}
\end{eqnarray}
together with relations that follow from these by complex conjugation.
Here the $\g_\m$'s are Dirac matrices of $SO(1,4)$, and $\g_{\m \n} = \ft
1 2 (\g_\m \g_\n - \g_\n \g_\m)$.  All other commutators vanish.

The list of generators we have given here constitutes the {\sl
superconformal decomposition} of the $SU(2,2|4)$ superalgebra.  This
is the form of the algebra most convenient for the study of conformal
superspace, though it is not as well adapted to physics in $AdS_5
\times S^5$.  For example, the generators of translations in
the $X^{\m}$ directions of $AdS_5$ are not $P_{\m}$, but rather the
linear combinations $\ft 1 2 (P_{\m} + K_{\m})$.

\end{document}